\documentclass[traditabstract]{aa} 

\usepackage{graphicx}
\usepackage{txfonts}

\usepackage{natbib}
\bibpunct{(}{)}{;}{a}{}{,} 
\usepackage[colorlinks=true,linkcolor=blue, citecolor=blue, urlcolor=blue]{hyperref}%

\usepackage{amssymb}	
\usepackage[dvipsnames]{xcolor} 
\usepackage{color,soul} 

\usepackage{lscape}

\begin{document} 

   \title{A Virgo Environmental Survey Tracing Ionised Gas Emission (VESTIGE). XI. Two dimensional H$\alpha$ kinematics 
   of the edge-on ram pressure stripped galaxy NGC\,4330\thanks{
   Based on observations obtained with
   (1) at the Observatoire de Haute Provence (OHP, France), operated by the French CNRS; 
   (2) at the Observatorio Astron\'omico Nacional at San Pedro M\'artir, Baja California, M\'exico (OAN -SPM) and with
   (3) MegaPrime/MegaCam, a joint project of CFHT and CEA/DAPNIA, at the Canada-French-Hawaii Telescope (CFHT) which is operated by the National Research Council (NRC) of Canada, the Institut National
   des Sciences de l'Univers of the Centre National de la Recherche Scientifique (CNRS) of France and
   the University of Hawaii. }
      }
   \subtitle{}
  \author{M. M. Sardaneta\inst{1},
P. Amram\inst{1},
A. Boselli\inst{1},
B. Vollmer\inst{2},
M. Rosado\inst{3},
M. S\'anchez-Cruces\inst{1,3},
A. Longobardi\inst{4},
C. Adami\inst{1},
M. Fossati\inst{4}, 
B. Epinat\inst{1},
M. Boquien\inst{5},
P. C{\^o}t{\'e}\inst{6},
G. Hensler\inst{7}, 
Junais\inst{1},
H. Plana\inst{8},
J.C. Cuillandre\inst{9},
L. Ferrarese\inst{6},
J.L. Gach\inst{1},
J. A. Gomez-Lopez\inst{1},
S. Gwyn\inst{6},
G. Trinchieri\inst{10} 
      }

\institute{     
        Aix-Marseille Univ., CNRS, CNES, LAM, Marseille, France 
        \email{minerva.munoz@lam.fr, philippe.amram@lam.fr, alessandro.boselli@lam.fr}
	\and
	Universit\'e de Strasbourg, CNRS, Observatoire astronomique de Strasbourg, UMR 7550, F-67000 Strasbourg, France
	\and
	Instituto de Astronom\'ia, Universidad Nacional Aut\'onoma de M\'exico, Apartado Postal 70-264, Coyoac\'an, Ciudad de M\'exico, CP 04510
        \and
	Universit\'a di Milano-Bicocca, piazza della scienza 3, 20100 Milano, Italy
	\and
	Centro de Astronom\'a (CITEVA), Universidad de Antofagasta, Avenida Angamos 601, Antofagasta, Chile
	\and
	National Research Council of Canada, Herzberg Astronomy and Astrophysics, 5071 West Saanich Road, Victoria, BC, V9E 2E7, Canada
	\and
	Department of Astrophysics, University of Vienna, T\"urkenschanzstrasse 17, 1180 Vienna, Austria
	\and
  	Laborat\'orio de Astrof\'isica Te\'orica e Observacional, Universidade Estadual de Santa Cruz - 45650-000, Ilh\'eus-BA, Brasil
	\and
	AIM, CEA, CNRS, Universit\'e Paris-Saclay, Universit\'e Paris Diderot, Sorbonne Paris Cit\'e, Observatoire de Paris, PSL University, F-91191 Gif-sur-Yvette Cedex, France
	\and
	INAF - Osservatorio Astronomico di Brera, via Brera 28, 20159 Milano, Italy
	}    
%

\authorrunning{Sardaneta et al.}
\titlerunning{VESTIGE}

\date{\today}

\abstract
  {Using the VESTIGE survey, a deep narrow-band H$\alpha$ imaging survey of the Virgo cluster carried on at the CFHT with MegaCam, we have discovered 
  a long and diffuse tail of ionised gas in the edge-on late-type galaxy NGC\,4330. This peculiar feature witnesses an ongoing ram pressure stripping event able to remove
  the gas in the outer region of the disc. Tuned hydrodynamic simulations suggest that the ram pressure stripping event is occurring almost face-on, making NGC\,4330
  the ideal candidate to study the effects of the perturbation in the direction perpendicular to the disc plane. We present
  here two new independent sets of Fabry-Perot 
  observations ($R$\,$\simeq$\,10000) with the purpose of understanding the effects of ram pressure stripping process on the kinematics of the ionised gas. Despite their 
  limited sensitivity to the diffuse gas emission, the data allowed us to measure the velocity and the velocity dispersion fields over the galaxy disc 
  and in several features at the edges or outside the stellar disc formed after the ram pressure stripping event. We have constructed the
  position-velocity diagrams and the rotation curves of the galaxy using three different techniques. The data show, consistent with the hydrodynamic simulations, that the galaxy has
  an inner solid-body rotation up to $\sim$\,2.4\,kpc, with non-circular streaming motions outwards the disc and in the several external features formed during the interaction
  of the galaxy with the surrounding intracluster medium. The data also indicate a decrease of the rotational velocity of the gas with increasing distance from
  the galaxy disc along the tails, suggesting a gradual but not linear loss of angular momentum in the stripped gas.
  Consistent with a ram pressure stripping scenario, the $i$-band image shows a boxy shape at the southwest edge of the disc, where the stellar orbits might have been perturbed by the modification of the gravitational potential well of the galaxy due to the displacement of the gas in the $z$-direction.
}


\maketitle
%

\section{Introduction}

The environment plays a major role in shaping galaxy evolution \citep[e.g.][]{Peng-2010ApJ...721..193P}. Rich clusters are dominated by quiescent 
systems \citep[morphology-segregation effect,][]{dressler-1980}, and the few spirals which inhabit these environments are generally deficient in 
all the components of their interstellar medium
(HI - \citeauthor{cayatte-1990}\,\citeyear{cayatte-1990}, \citeauthor{Solanes-2001ApJ...548...97S}\,\citeyear{Solanes-2001ApJ...548...97S}, 
\citeauthor{gavazzi-2005A&A...429..439G}\,\citeyear{gavazzi-2005A&A...429..439G}, \citeauthor{chung-2009}\,\citeyear{chung-2009};
H$_2$ - \citeauthor{Fumagalli-2009}\,\citeyear{Fumagalli-2009}, \citeauthor{boselli-2014}\,\citeyear{boselli-2014};
dust - \citeauthor{cortese-2010-a}\,\citeyear{cortese-2010-a}, \citeauthor{cortese-2012-a}\,\citeyear{cortese-2012-a},
\citeauthor{longobardi-2020}\,\citeyear{longobardi-2020}).
Their star formation is also reduced with respect to that of similar objects in the field 
\citep{gavazzi-1998, gavazzi-2002-b, gavazzi-2002-a, gavazzi-2006A&A...446..839G,  gavazzi-2010A&A...517A..73G, gavazzi-2013A&A...553A..89G, 
Lewis-2002MNRAS.334..673L, gomez-2003ApJ...584..210G, Peng-2010ApJ...721..193P, boselli-2014-ii, boselli-2016-a, cortese-2021}.

Different physical mechanisms have been proposed in the literature to explain these results, as extensively reviewed in \citet{boselli-2006-a, boselli-2014-iv}.
These include gravitational interactions with other cluster members \citep{Merritt-1983ApJ...264...24M} or with the gravitational potential well of the 
cluster \citep{byrd-1990}, including their combined effect, generally called galaxy harassment \citep{Moore-1998ApJ...495..139M}, or the hydrodynamic 
interaction of the interstellar medium (ISM) of galaxies with the hot 
and dense intracluster gas (thermal evaporation, \citeauthor{cowie-1977-Natur}\,\citeyear{cowie-1977-Natur}; starvation, 
\citeauthor{Larson-1980ApJ...237..692L}\,\citeyear{Larson-1980ApJ...237..692L}; ram pressure stripping, \citeauthor{gunn-1972}\,\citeyear{gunn-1972}). 
Several observations seem to indicate that while gravitational perturbations are dominant in relatively small systems such as groups thanks to the low 
velocity dispersion of galaxies within the dark matter halo, hydrodynamic 
interactions are dominant in massive clusters, where the high velocity dispersion and the density of the gas make them particularly efficient 
\citep[e.g.][]{vollmer-2001, boselli-2008-a, boselli-2008-b, boselli-2014, gavazzi-2013A&A...553A..89G}. It is nevertheless clear that all these 
different perturbing mechanisms might be acting jointly on galaxies in rich environments, with different effects on their evolution.

Despite this evolutionary picture is becoming clearer, there are still several aspects of the galaxy transformation in rich environments which are 
still poorly known. Among these is the fate of the ISM removed during a ram pressure stripping event. The stripped ISM is often observed in spectacular 
tails of atomic \citep[e.g.][]{chung-2007}, ionised \citep[e.g.][]{gavazzi-2001, Yagi-2010AJ....140.1814Y, boselli-2016-b}, or hot gas 
\citep[e.g.][]{Sun-2007ApJ...671..190S}. Indeed, it has been observed that while in some objects the stripped gas collapses within the tail 
to form giant molecular clouds 
\citep[][]{jachym-2014, jachym-2017ApJ...839..114J, jachym-2019ApJ...883..145J, Moretti-2020ApJ...897L..30M}
 and stars \citep[][]{Fossati-2016MNRAS.455.2028F, Poggianti-2019MNRAS.482.4466P}, in other systems it remains diffuse \citep[e.g.][]{Boissier-2012}. 
In these objects the cold gas stripped from the disc does not form stars because it changes of phase becoming first ionised than hot \cite[e.g.][]{boselli-2016-b}. The detailed 
 analysis of a few nearby systems with high quality multifrequency spectroscopic and imaging data seem to suggest that the gas collapses into giant 
 molecular clouds only whenever its velocity dispersion is sufficiently low 
\citep[e.g.][]{Fossati-2016MNRAS.455.2028F, boselli-2018-ii}. Otherwise, the cold gas changes of phase for heat conduction, turbulence, 
magneto-hydrodynamic waves or shocks. It thus becomes first ionised and then hot gas visible in the X-rays spectral domain 
\citep{gavazzi-2001, Sun-2007ApJ...671..190S, Fossati-2016MNRAS.455.2028F, boselli-2016-b}.

The kinematics of the stripped gas is thus a crucial parameter for understanding the process of star formation within the tails.
Observations and simulations of ram pressure stripped galaxies consistently 
indicate that a ram pressure stripping process is able to drastically perturb the velocity field of a spiral galaxy, displacing its kinematic centre  
with respect to the photometric one, and affecting its rotation curve \citep{vollmer-2001, vollmer-2004-a, vollmer-2006, vollmer-2008, 
Kronberger-2008-a, Merluzzi-2013MNRAS.429.1747M, Fumagalli-2014, Consolandi-2017, Sheen-2017ApJ...840L...7S, bellhouse-2017, bellhouse-2019, boselli-2021}. 
Despite the advent of extraordinary new instrumentation in the optical 
(MUSE at the VLT) and radio (ALMA) domain, multifrequency data at high angular and spectral resolution of nearby ram pressure stripped galaxies, 
where the proximity allows us to study the star formation process down to the scale of individual HII regions, are only sporadically available. 
To understand the effects of the stripping process on the kinematics of the gas, observations and simulations of galaxies spanning a wide range 
in stellar mass and with different impact parameters are necessary. A particularly interesting case are edge-on galaxies which allow us to 
investigate as the gas is stripped out from the disc plane in the $z$-direction. Among these, there are only a few systems undergoing a ram pressure 
stripping event which have been studied in detail: UGC\,6697 in the cluster A1367 \citep{Consolandi-2017}, and JO204 in the further cluster A957 \citep[$z \sim$\,0.052,][]{Gullieuszik-2017ApJ...846...27G}. The kinematic properties of the stripped gas are significantly different in the two galaxies given their impact parameters, UGC\,6697 
suffering an edge-on stripping event while JO204 a face-on one.

The Virgo Environmental Survey Tracing Ionised Gas Emission \citep[VESTIGE,][]{boselli-2018-i}, a blind H$\alpha$ narrow-band imaging survey of Virgo, 
allows us to identify a large number of galaxies undergoing a ram pressure stripping event in the closest cluster of galaxies. The presence of ionised 
gas tails without any stellar counterpart is the clearest demonstration that a system is undergoing a hydrodynamic interaction with the surrounding 
environment. Indeed, gravitational perturbations act indifferently on the stellar and gaseous components, also producing tidal structures visible in 
deep broad-band imaging data. Several galaxies undergoing a ram pressure stripping event have been identified so far 
\citep{boselli-2016-b, boselli-2018-iii, boselli-2021, fossati-2018, longobardi-2020, junais-2021}. Among them, NGC\,4330 is an edge-on galaxy 
with evident tails of atomic \citep{chung-2007, chung-2009} and ionised \citep{abramson-2011, fossati-2018} gas indicating that the galaxy is 
entering the cluster from the south and impacting the intra cluster medium almost face on. It is thus an ideal candidate to study in great detail 
the effects of the perturbation on the kinematics of the perturbed gas on the disc of the galaxy and within the tail.

In this paper we use two 
different sets of high spectral resolution (up to $R$\,$\simeq$\,10000) Fabry-Perot (FP) data to study the kinematics of the ionised gas in this 
representative object. We combine these new sets of observations with tuned hydrodynamic simulations developed to reproduce the structural and 
spectrophotometric properties of this galaxy \citep{vollmer-2021}. Unfortunately most of the gas in the tail has a very low surface brightness, 
and thus it is out of reach of our dedicated observations which detect only a few, high surface brightness star forming regions. The proximity 
of the galaxy \citep[16.5\,Mpc,][]{gavazzi-1999, mei-2007}, however, and the adopted spectral resolution allow us to study in detail the kinematics 
of the gas of these compact regions down to scales of $\sim$100\,pc in angular resolution and velocity dispersion $\sigma \sim 13 \mathrm{km\ s^{-1}}$. 
Such resolution unreached in the more distant UGC\,6697 and JO204 despite their extraordinary MUSE data.

The paper is structured as follows: in Sect. \ref{section2} we describe the MegaCam CFHT narrow-band imaging and the FP high-resolution spectroscopic 
observations along with the multifrequency data used in the analysis. In Sect. \ref{section3} and \ref{section4} we analyse the imaging data and construct the rotation 
curve of the galaxy, while in Sect. \ref{section5} we study the 2-D kinematics of the gas and the specific kinematics of the gas out of the plane in Sect. \ref{section6}. The results of 
the analysis are compared to the hydrodynamic simulations of \citet{vollmer-2021} in Sect. \ref{section7}. The discussion and conclusions are given in Sect. \ref{section8}.

\section{Observations and data reduction}
\label{section2}

\begin{table}
\caption{General properties of NGC\,4330}
\label{general}
\begin{center}
\begin{tabular}{lcc}
\hline\hline 
Variable 				& Value 	& Ref. 	\\ 
\hline
Alternative names  $^{(1)}$					& HRS 124				& a \\ 
$\alpha$ (J2000) $^{(2)}$					& $12^h 23^m 17.10^s$   & b  \\ 
$\delta$ (J2000) $^{(3)}$					& $+11^{\circ} 22' 5''.7$  	& b  \\ 
Type 						& Scd? 									& a \\ 
Distance  					& 16.5\,Mpc 								& c \\ 
Dist. from M87 $^{(4)}$	& 2.1$^{\circ}$\,/\,604.8\,kpc 								& TW \\ 
$D_{25}$     & 5.86\,arcmin\,/\,28.1\,kpc & d\\
$i_{D_{25}}$  				& 11.925\,$mag$								& d \\ 
$V_{heliocentric}$  		& 1551\,km\,s$^{-1}$ 									& TW \\ 
$V_{rot}^{max}$  			& 140\,km\,s$^{-1}$ 									& e \\ 
$PA$  				& 59$\pm$1$^{\circ}$ 								& TW  \\ 
Morphological Inclination  	& 89.5$^{\circ}$						& TW \\ 
$M_{star}$   			& 5 $\times$ 10$^9$	M$_{\odot}$			& f \\
$M(HI)$  				& 4.73 $\pm$ 1.04 $\times$ 10$^8$ M$_{\odot}$ 			& e \\ 
$M(H_2)$  				& 1.27 $\pm$ 0.05 $\times$ 10$^8$ M$_{\odot}$ 		& g \\ 
\hline\\
\end{tabular}
\end{center} 
\tablefoot{
$^{(1)}$ UGC 7456, VCC 630, IRAS 12207+1138. 
$^{(2)}$ Right ascension and $^{(3)}$declination of the galaxy. 
$^{(4)}$ Projected distance for M\,87. %
}
\tablebib{TW: This work; a: NED; b: NGVS, \cite{ferrarese-2012}; c: \cite{gavazzi-1999, mei-2007}; 
d: \cite{cortese-2012-b};
e: \cite{chung-2009}; f: \cite{boselli-2015}; g: \cite{lee-2017}.
}
\end{table}


\subsection{VESTIGE narrow-band imaging}

Deep H$\alpha$ narrow-band imaging of NGC\,4330 has been gathered with MegaCam at the CFHT by the VESTIGE survey \citep{boselli-2018-i},
and the first imaging results have been presented in  \citet{fossati-2018}.
Briefly, the galaxy has been observed using the narrow-band filter MP9603 ($\lambda_c$\,=\,6591\,\AA; $\Delta \lambda$\,=\,106\,\AA), which includes 
the H$\alpha$ line and the two [NII] lines at $\lambda$\,6548\,\AA\ and $\lambda$\,6584\,\AA 
\footnote{Hereafter we refer to the H$\alpha$+[NII] band simply as H$\alpha$, unless otherwise stated.}. 
The galaxy has been observed during the main VESTIGE survey but also during a pilot run. 
The data used in this work are the combination of these two datasets, and are thus deeper than
those analysed in \citet{fossati-2018}, which were limited to the pilot observations. To secure the subtraction of the stellar continuum, the galaxy was 
also observed in the broad $r$-band filter with shorter exposures to avoid saturation within the nucleus. The different exposures were gathered using a 
large dither pattern (15 arcmin in RA and 20 arcmin in Dec) to minimise the effects of scattered light and reflections in the flat-fielding, thus optimising 
the detection of low surface brightness features in the ionised gas component. A slightly different pattern was used for the
pilot observations. The total integration 
time is 21060 sec in the narrow-band filter and 1506 sec in the broad-band $r$ filter\footnote{The total integration time 
in the $r$-band filter reported in \citet{fossati-2018} includes archival data not considered here for consistency with the rest of the VESTIGE survey.}. 
These deep observations have a typical sensitivity of 
$f(\mathrm{H}\alpha) \simeq 2\times 10^{-17}$\,erg\,s$^{-1}$\,cm$^{-2}$ (5\,$\sigma$) for point sources 
and $\Sigma(\mathrm{H}\alpha)$ $\simeq$10$^{-18}$erg\,s$^{-1}$cm$^{-2}$\,arcsec$^{-2}$ (1\,$\sigma$ after smoothing the data to $\sim$\,3 
arcsec resolution) for extended sources, thus approximately a factor of 1.7 deeper than the mean sensitivity of VESTIGE. NGC\,4330 has been observed during excellent seeing conditions (FWHM\,=\,0.73\,arcsec).

The data have been reduced using Elixir-LSB \citep{ferrarese-2012}, a data reduction pipeline especially designed to detect the diffuse emission of 
extended low surface brightness structures formed during the interaction of galaxies with their surrounding environment \citep[e.g.][]{boselli-2016-a}. 
The astrometric and photometric calibrations have been done using the SDSS and the PanSTARRS surveys by means of the MegaPipe pipeline 
\citep{Gwyn-2008PASP..120..212G}. The typical photometric uncertainty in the data is of 0.02-0.03 mag in both bands.
The emission of the ionised gas has been derived after subtracting the stellar continuum as described in \cite{boselli-2019}. This is done by
combining the $r$-band image with the $g$-band frame obtained during the Next Generation Virgo Cluster Survey \citep[NGVS,][]{ferrarese-2012} to take into account the colour 
of the stellar continuum \citep{Spector-2012MNRAS.419.2156S}.

\subsection{Fabry-Perot spectroscopy}

High resolution IFU spectroscopy was obtained using two different FP interferometers, PUMA attached at the 2.1 m telescope at San Pedro M\'artir, Baja California, M\'exico, and GHASP, mounted on the 1.93 m telescope at the Observatoire de Haute Provence, France. A detailed description of the instruments, of the observations, and of the data reduction is given in Appendix \ref{FabryPerotObservations}. Briefly, PUMA is a FP spectrograph with a field of view of $\sim$10$\times$10\,arcmin$^2$ coupled with a 512$\times$512 CCD camera, with pixel size of 1.27$\times$1.27\,arcsec$^2$ and has a spectral resolution $R\sim$\,6500. 
GHASP has a field of view of 5.8$\times$5.8\,arcmin$^2$, is coupled with a 512$\times$512 Image Photon Counting System with pixel size of 0.68$\times$0.68\,arcsec$^2$, and has a spectral resolution $R\sim$10000.

\subsection{Multifrequency data}\label{Subsect.Multifreq}

The FP data were combined to those gathered during other surveys of the Virgo cluster for the following analysis, as extensively described in 
\cite{fossati-2018}. Namely, the distribution of the old stellar population has been derived using the very deep $i$-band image obtained during the NGVS 
survey of the cluster \citep{ferrarese-2012}. HI data are also available from the Imaging of Virgo Spirals in Atomic Gas (VIVA), gathered at the VLA in 
C-short array configuration \citep{chung-2009}. These data have been obtained  with a velocity sampling of $\sim\mathrm{10\,km\,s^{-1}}$ and a spatial resolution 
of $\sim$15\,arcsec. The same set of data has been combined with other VLA data obtained in D configuration to increase the sensitivity to low column densitiy, 
extended features by \cite{abramson-2011} with a final velocity resolution of 10.4\,km\,s$^{-1}$ and an angular resolution of 
FWHM of 26.4$\times$24.0\,arcsec$^{2}$.
$^{12}\mathrm{CO}\ (2-1)$ observations obtained with the Submillimeter Array (SMA), with a synthesized beam of 6.35$\times$4.47\,arcsec$^{2}$ and a spectral resolution 
of 5\,km\,s$^{-1}$, have been presented in \cite{lee-2017}.

\section{Imaging analysis }
\label{section3}

\begin{figure*}
\centering
\includegraphics[width=1\linewidth]{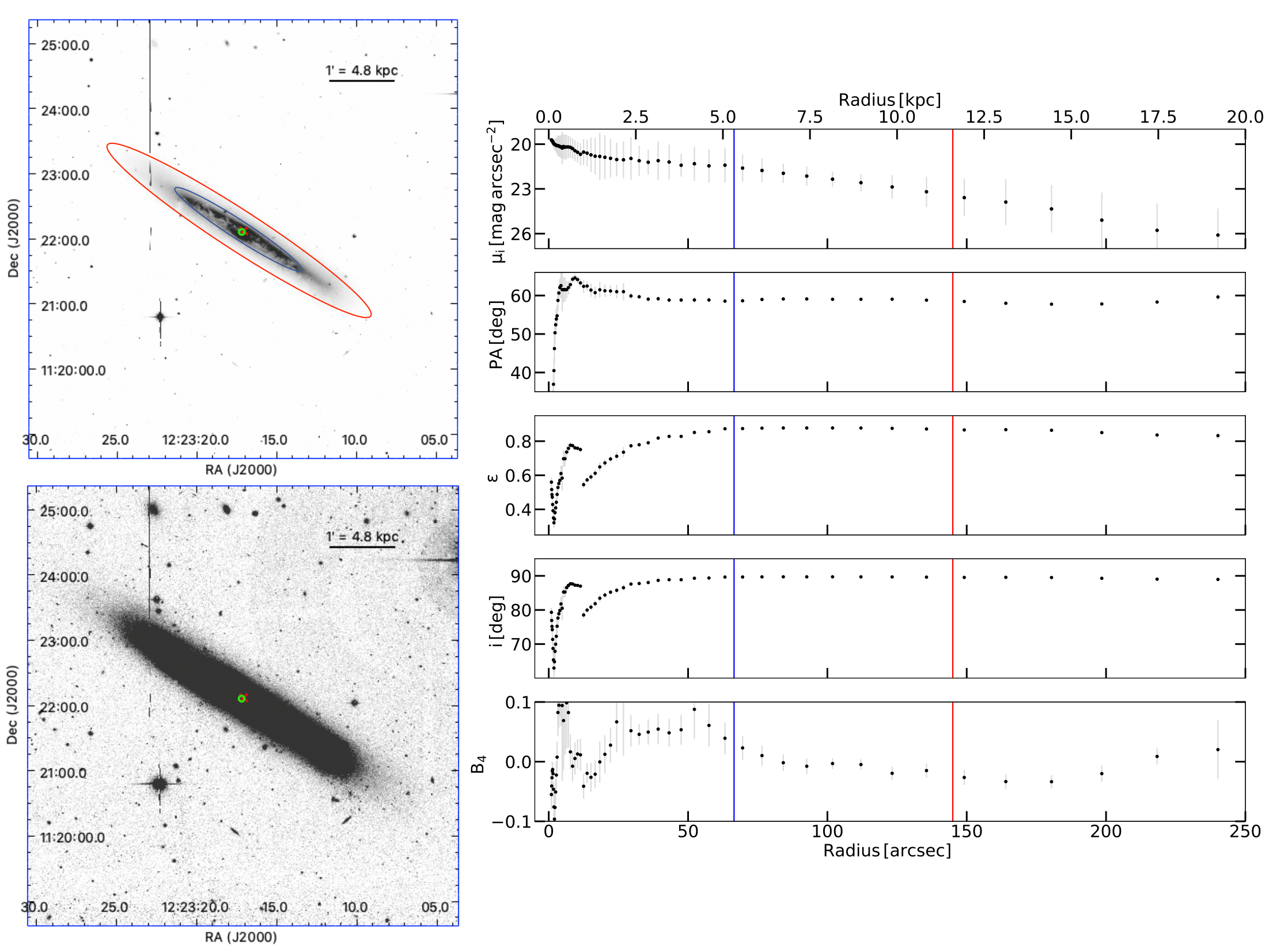}
    \caption{Left panels: NGVS $i$-band image of NGC\,4330 at low contrast (upper panel) to see the distribution of the absorbing dust within the disc, and at high contrast in logarithmic scale 
    (lower panel) to see the boxy shape of the disc at its edges. The red  and blue ellipses indicate the $i$-band mean surface brightness
    $\mu(i)$\,=\,23.5 and the effective surface brightness $\mu_{e}$=\,21.5\,$\mathrm{\,mag\,arcsec^{-2}}$, respectively. The green bullet and the red cross indicate the position of the 
    photometric and kinematical centre, respectively. Right panels, from top to bottom, radial variation of the $i$-band surface brightness ($\mu$), position angle ($PA$), ellipticity ($\varepsilon$), morphological inclination ($i_{ph}$), and the fourth degree cosine coefficient of the Fourier series ($B_4$) parameters. The red and blue vertical lines correspond to the ellipses drawn on the upper left panel and indicate the radius $R_{23.5}(i)\sim$145.0\,arcsec ($\sim$11.6\,kpc) and the effective radius $R_{e}(i)\sim$66.5\,arcsec ($\sim$5.3\,kpc), respectively.}
    \label{iband}
\end{figure*}
    
\subsection{Stellar distribution}

\begin{figure*}
\begin{center}
\includegraphics[width=0.8\linewidth]{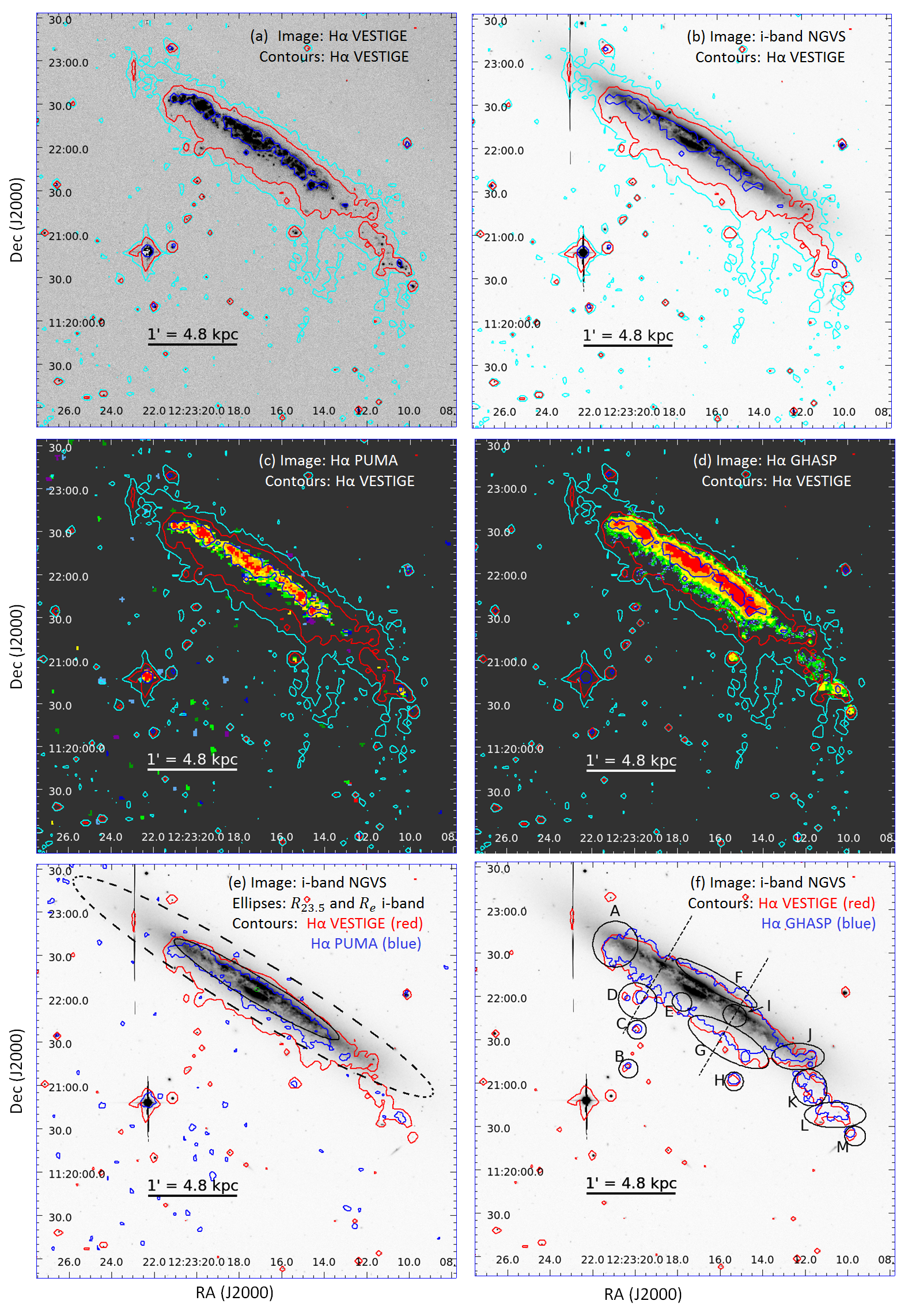}
    \caption{
    Images of NGC\,4330.
    Panel  (a):  full resolution VESTIGE continuum-subtracted H$\alpha$ narrow-band image of NGC\,4330 in linear scale.
    Panel (b): NGVS $i$-band image.
    Panels (c) and (d): PUMA and GHASP FP H$\alpha$ monochromatic images, respectively.
    Panel (e): NGVS $i$-band image, the dashed black ellipse and the inner ellipse indicate the mean  surface brightness, $\mu(i)$=$23.5$,  and the effective surface brightness, $\mu_{e}(i)$=$21.5\,\mathrm{\,mag\,arcsec^{-2}}$, respectively, the green bullet indicates the photometric centre (see Fig. \ref{iband}). 
    Panel (f): NGVS $i$-band image with marks on the zones of interest labelled by letters (A to M) and two parallel dashed lines at  $r_{0}\sim30$\,arcsec ($\sim$2.4\,kpc) which divide the galactic disc in three main regions: northeastern, central and southwestern (see Sect. \ref{section4}). 
The cyan, red, and blue contours in panels (a) to (d) show the VESTIGE surface brightness $\Sigma(\mathrm{H}\alpha)$=0.3,\,1.0\,and\,12.5$\times$10$^{-17}$ $\mathrm{erg\ s^{-1}\ cm^{-2}\, arcsec^{-2}}$, respectively.
The red contours in panel (e) and (f) show the VESTIGE surface brightness $\Sigma(H\alpha)$=10$^{-17}$ $\mathrm{erg\ s^{-1}\ cm^{-2}\, arcsec^{-2}}$.
The blue contours in the lowest left (e) panel indicate the surface brightness limit of PUMA $\Sigma(H\alpha)\sim$3.5$\times$10$^{-17}\,\mathrm{erg\ s^{-1}\ cm^{-2}\, arcsec^{-2}}$, while those
in panel (f) show the surface brightness limit of GHASP $\Sigma(H\alpha)\sim$10$^{-17}\,\mathrm{erg\ s^{-1}\ cm^{-2}\, arcsec^{-2}}$.
}
\label{Hacomposite}
\end{center}
\end{figure*}

The NGVS $i$-band image (see Fig. \ref{iband}) traces the bulk of the stellar population and can 
thus be used to derive the main morphological and structural parameters of the galaxy. It can also be used as reference to 
identify the different ionised gas structures removed from the disc plane during the interaction. This image shows 
an edge-on disc with a symmetric and unperturbed stellar distribution down to the surface brightness limit of 
the NGVS survey ($\mu_{i}$\,=\,27.4\,mag\,arcsec$^{-2}$), proving that the galaxy is not undergoing any major 
gravitational perturbation. The galaxy has a small bulge without any evident bar \citep[e.g.][]{abramson-2011}.

We derived the photometric properties of the stellar disc through the elliptical isophote fitting method 
using the IRAF task \textsc{ellipse} \citep{Jedrzejewski-1987}. In this task,  
ellipses are drawn to match the isophotes. The surface brightness along the ellipse is expanded in Fourier series:
\begin{equation}
\mu(\theta) = \mu_0+ \Sigma_{n= 1,\infty}\ (A_n \sin n \theta + B_n \cos n \theta),
\label{ellipse}
\end{equation}
where $\mu_0$ is the surface brightness averaged over the ellipse, $A_n$ and $B_n$ the higher order Fourier 
coefficients. If an isophote is a perfect ellipse, all the $A_n$ and $B_n$ coefficient would be exactly zero. 
The term $A_1$, $B_1$, $A_2$, and $B_2$ indicate errors in the fitting procedure.  The terms $A_3$ and $B_3$ 
give the ``egg-shaped" isophote but the most interesting is $B_4$. If $B_4>0$, the galaxy has discy-shape isophotes, 
while if $B_4<0$ boxy-shaped isophotes.  The \textsc{ellipse} task provides a number of
 parameters that describe the surface brightness of the galaxy as a function of the semi-major axis, including 
 the ellipticity ($\varepsilon$), the position angle ($PA$), and it reconstructs the growth curve which gives 
 the total apparent magnitude inside each isophote \citep[e.g.][]{barbosa-2015, korsaga-2019}.  
All these parameters were allowed to vary as a function of the radial distance from the galaxy centre.

The isophotal radius derived using the NGVS $i$-band image is $R_{23.5}(i)\sim$145.0\,arcsec ($\sim$11.6\,kpc). This value is slightly larger than 
the one derived by \cite{cortese-2012-b} using the SDSS-$i$-band image ($R_{23.5}(i)\sim$140.3\,arcsec, $\sim$11.2\,kpc).  
The difference with this previous measurement might be due to the presence of dust and to the adopted sigma clipping selection in the \textsc{ellipse} task. 
The effective radius is $R_{e}(i)\sim$66.5\,arcsec ($\sim$5\,kpc) and the effective surface brightness $\mu_{e}(i)\sim$21.5$\,\mathrm{\,mag\,\,\,arcsec^{-2}}$, 
consistently with those given in \cite{cortese-2012-b}. 
The photometric centre is located at $\mathrm{RA(J2000)}=12^{h}23^{m}17.26^{s}$, $\mathrm{Dec}=+11^{\circ} 22' 07''.9$, with an uncertainty of $\sim$0.9\,arcsec\,($\sim$\,76 pc). 
The sudden jump of the parameters observed in Fig. \ref{iband} at a radius of $\sim$\,5\,arcsec ($\sim$0.4\,kpc) is due to the dusty region around the centre.

Consistently with e.g. \cite{barbosa-2015} we estimate the morphological inclination of the galaxy using the relation:
\begin{equation}
\cos i_{ph} = \frac{(1-\varepsilon)^{2}-q_{0}^{2}}{1-q_{0}^{2}},
\label{inclination}
\end{equation}
where $q_0$\,=\,0.1 is the intrinsic flattening of a Scd galaxy \citep{haynes-1984}.  
Figure \ref{iband} indicates that the morphological inclination of the galaxy converges to $i_{ph} =  89.5 ^{\circ}$ and the photometric 
position angle to $PA_{ph}= 59  \pm 1^{\circ}$ for $R(i)$\,$\geq$\,75\,arcsec\,($\sim$6\,kpc) ($\mu_{i}$\,$\geq$\,22.75\,mag\,arcsec$^{-2}$).  
The $B_4$ parameter, which is negative, indicates that the galaxy disc has a boxy shape outside of the effective radius, $R_{e}$ shown in Fig.\,\ref{iband}. This particular shape becomes more evident in the saturated $i$-band image 
and is significantly stronger in the southwest direction. Tuned simulations suggest that this particular 
shape could be produced after a ram pressure stripping event. The gas removed out from the disc plane in the $z$-direction can alter the shape 
of the gravitational potential well, modifying the stellar orbits, thus producing a thicker disc
\citep[][]{Farouki-1980, clarke-2017, Safarzadeh-2017ApJ...850...99S, Steyrleithner-2020MNRAS}. This phenomenon might explain the peculiar morphology of NGC\,4330, with a boxy shape 
particularly evident in the southwest direction, where the stripping is more efficient. Unfortunately, we lack of stellar kinematical data 
necessary to further probe this scenario.

\subsection{Warm gas distribution}

The ionised gas distribution of NGC\,4330 has been extensively described in \cite{fossati-2018}. Here we summarise the main characteristics, 
referring the reader to that work for a more detailed description. The ionised gas emission mainly comes from the inner regions of the stellar disc, where
most of the HII regions are situated (see Fig. \ref{Hacomposite}). The H$\alpha$ disc is truncated with respect to the stellar disc, indicating that
the activity of star formation has been quenched at its edges \citep{fossati-2018}. The northeastern edge of the star forming disc is bent to the south in a "hook" 
shaped structure \citep[e.g.][]{abramson-2011}, while a few compact and bright HII regions are present just south to the stellar disc at the southwestern edge. 
There is also a very extended,
diffuse and low surface brightness emission ($\Sigma(H\alpha)\sim$\,3$\times$10$^{-18}$\,erg\,s$^{-1}$\,cm$^{-2}$\,arcsec$^{-2}$) towards the south of the disc, produced by the gas stripped during the ram pressure episode, indicating that the galaxy
is moving north-worth on the plane of the sky within the cluster. Associated to the diffuse emission, the VESTIGE H$\alpha$ image shows a few weak compact regions with a surface brightness over 
$\Sigma(H\alpha)$ $\sim$\,10$^{-17}$\,erg\,s$^{-1}$\,cm$^{-2}$\,arcsec$^{-2}$, which corresponds to the GHASP detection limit where star formation is probably taking place.
Not all the features located out of the plane of the galaxy and above this threshold limit are clearly detected in the monochromatic FP 
image of NGC\,4330, neither in the PUMA data nor in the deeper GHASP data. 
The bent region at the northeast edge of the disc (labelled A in Fig. \ref{Hacomposite}\,f) is clearly detected as well as a few regions on the southwest extension 
of the disc (labelled K, L and M).
A few ionised regions have also been detected at the limit of the stellar disc (labelled E, G and F).

\section{Rotation curves}
\label{section4}

\begin{figure*}[ht!]
\centering
\includegraphics[width=0.49\linewidth]{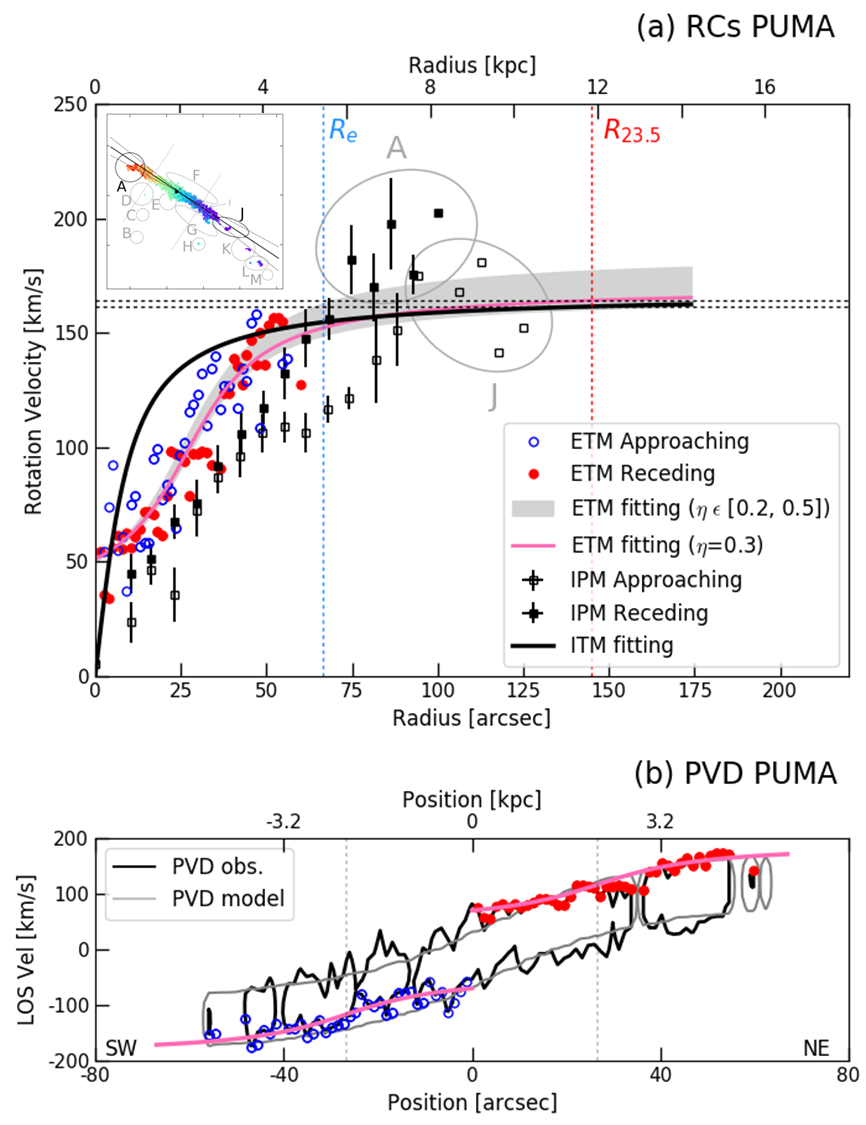}
\includegraphics[width=0.49\linewidth]{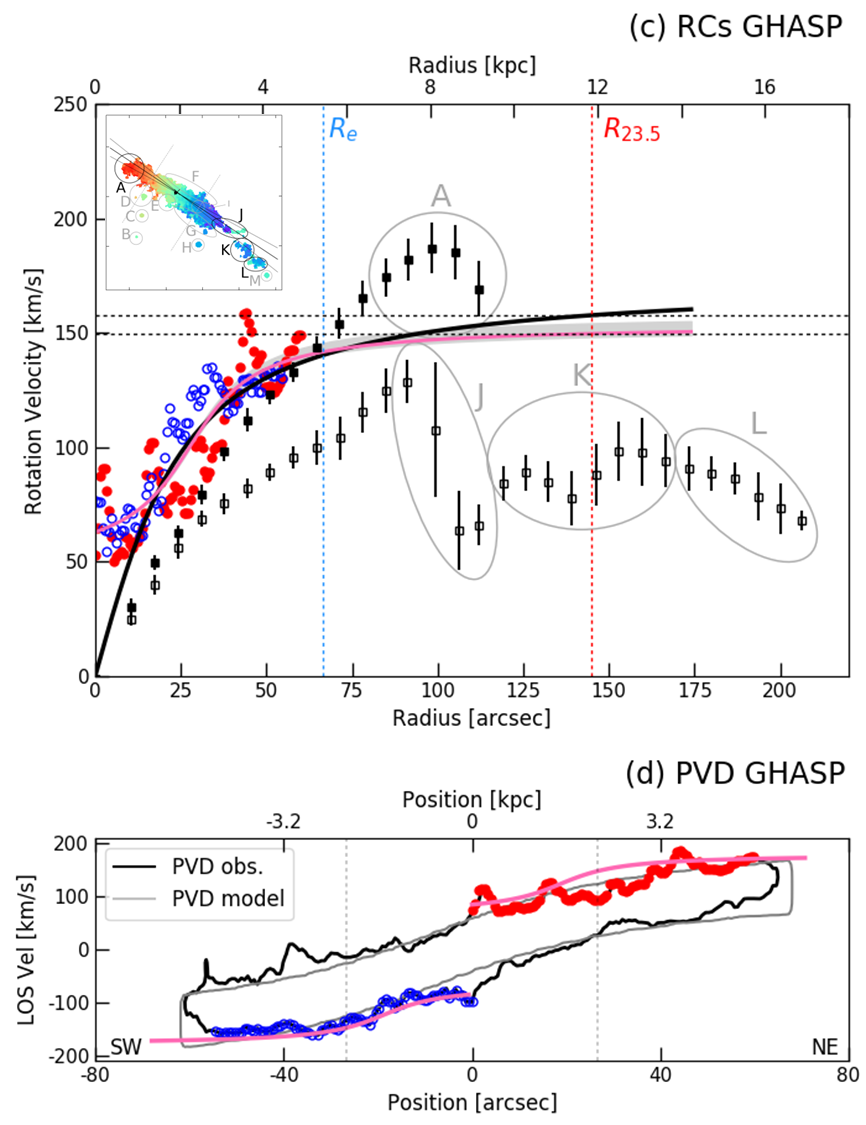}
\caption{PUMA (panels (a) and (b)) and GHASP (panels (c) and (d)) rotation curves (RCs, panels (a) and (c)) and position-velocity diagrams (PVDs, panels (b) and (d)) of 
NGC\,4330. Panels (a) and (c): RCs derived using three different methods:  the intensity-peak method (IPM), the envelope-tracing method (ETM) and 
the iteration-method (ITM) (see Sect. \ref{The envelope-tracing method} for a further explanation on the methodology). In the case of the IPM (black squares), an azimuthal sector of $\pm5^{\circ}$ in the sky plane has been taken around the major axis to compute the rotation velocities; filled and empty black squares indicate the receding and the approaching sides, respectively.
The upper-left insert shows the azimuthal sector and the location of the regions pointed on the rotation curve. For the ETM, the rotation velocities  have been derived from 
the PVD along the major axis ($z$=0), shown in panels (b) and (d).
The terminal velocities are represented by empty blue circles for the receding side and the filled red circles for the approaching side;
the pink line corresponds to the best fit to these points using $\eta=0.3$ and the shaded area shows the rotation velocity amplitude for $\eta=[0.2, 0.5]$ (see relation \ref{sofue-ec-intensity}).
The ITM based on a cylindrical model (described in section \ref{ITM}) is represented by the black line.
Panels (b) and (d): position-velocity diagram (PVD) showing the LoS velocities computed with the ETM from which the rotation velocities are computed in the upper panels.
On top of the PVD, the filled red and open blue circles indicate the extremal LoS velocities and the pink curve the arctan model fitting those velocities.
The red and blue vertical dotted lines in panels (a) and (c), show the photometric and effective radius, the gray dotted lines in the lowest panels the position of $r_{0}$, see Sect. \ref{section4}.
}
   \label{RCET}
\end{figure*}

Rotation curves (RC) of galaxies are generally derived using the intensity-peak method (IPM). However, this method hardly works for galaxies with high inclination because the velocity profiles in a given position integrate the whole flux along the line of sight (LoS), thus at different radii on the galactic plane \citep[][]{sancisi-1979, swaters-1997, 
garcia-ruiz-2002, Rosado-2013}. To overcome this problem, two different methods are generally used: the envelope-tracing method (ETM)
\citep[][]{sofue-1997AJ....114.2428S, sofue-1999PASJ...51..737S, garcia-ruiz-2002} 
and the iteration method (ITM) \citep{swaters-1997, takamiya-2000, heald-2006, heald-2007}. 
We apply these three methods to NGC\,4330, following the methodology described in detail in Appendix \ref{DerivationOfTheRotationCurve}.

\subsection{Comparison of the three methods}

Figure \ref{RCET} shows the RCs derived using three different methods: (1) the intensity-peak method (IPM), (2) the envelope-tracing method (ETM), and (3) the iteration-method (ITM). 
The RCs obtained with the IPM using both GHASP and PUMA datasets show that NGC\,4330 is rotating as a solid-body inside a radius of $\sim$30\,arcsec.
The output of the fit of Eq. \ref{eq-vr-teorica} (ETM) provides similar terminal velocities reached around $\sim30$\,arcsec ($\sim$2.4\,kpc) as well as similar transition radii between the rising and the flat part of the RCs compared to the IPM method. 
After this transition, the RCs computed with the ETM method are flat up to the photometric radius $R_{23.5}\sim140$ arcsec ($\sim$11.6\,kpc), which is represented by the red line in panels (a) and (c) of Fig. \ref{RCET}. The approaching and receding sides of the RCs measured with the IPM method start to differ at a radius of $\sim 30$ arcsec ($\sim  2.5$\,kpc) and do not have a single point of matching where the two sides overlap until the photometric radius.
The rotation velocity obtained for each dataset with the ETM differs by 8 km s$^{-1}$. This number is within the uncertainties due to the differences in resolution between the two instruments and could be partly due to a different threshold for the PVD extraction.

Both datasets show that the RCs derived using the ETM are steeper than those derived using the IPM in the inner regions. 
This difference is due to the fact that here the contribution of regions along the same LoS but located at different radial distances from the galaxy centre is large, while this is not the case at the edges of the disc where the ETM measures the rotation. 
Close to the galaxy centre, the width of the PVD is supposed to narrow because along the minor axis, the rotation velocity vector lies in the sky plane, thus its projected component along the LoS is null. Since we do not observe a large decrease of the PVD width in the centre, we conclude that it is mainly due to radial velocity dispersion.

For the GHASP dataset, the RCs derived from the ETM and the ITM have a similar inflexion shape along the transition radius ($r_t$) between the rising and the flat part of the RC (see Fig. \ref{RCET}).  
Nevertheless, the fit shows that the parameter $r_t$ differs by $\sim$5\,arcsec ($\sim$0.4\,kpc) using the two methods. 
For the PUMA dataset, $r_t$ occurs at a radius $\sim$9\,arcsec ($\sim$0.7\,kpc)  larger with the ETM than with the ITM, which makes the RC steeper with this last method (see Fig. \ref{RCET}). Despite the difference between $r_t$ 
computed using the ETM and ITM, an acceptable discrepancy between the maximal rotational velocities of $\sim 8\,\,\,\mathrm{km\,s^{-1}}$ is observed at the photometric 
radius in the RCs of each dataset.

The three methods consistently show that the rotation of two-thirds of the ionised gas disc of NGC\,4330 (i.e. inside a diameter of $\sim$60 arcsec, $\sim$4.8\,kpc) 
can be modelled as a solid rotating body.
The same methods have also revealed that the transition between the rising and flat part of the RCs occurs in between $r_{0}\sim$30\,arcsec ($\sim$2.4\,kpc) and  
slightly before the photometric radius.

The IPM allows us to analyse the rotation on both the receding and approaching sides of the galaxy using the pixels within an angular sector of $\pm$5$^{\circ}$ 
from the kinematic major axis.
The RCs derived using this method show that at the photometric radius the rotation starts to be dominated by streaming motions. This occurs mainly at the southwest of the disc, where the gas has non-circular motions and where the ionised disc is distorted, hence the projection parameters may be no longer valid in these outer regions. These peculiar motions are not modelled with the other methods. Nevertheless, the ETM and ITM withdraw the 
effects of LoS to the rotational velocity on the inner disc, and allow us to obtain a reliable measure of the maximal rotational velocity of the galaxy, we fix it 
to the average value of the four velocities given in table \ref{pumavsghasp} for the two datasets: $V_{rot}^{max}=158\pm12$\,km\,s$^{-1}$, the uncertainty being estimated from their standard deviation.

\subsection{The dynamical mass}

The RC of NGC\,4330 is not flat because the gas which is used as a mass-tracer is subject to ram pressure stripping 
outside the transition radius. Indeed, taking into account its size, 
morphological type, and maximum rotation velocity,  NGC\,4330 is an intermediate-mass galaxy which is supposed to have a flat 
RC \citep[e.g.][]{sofue-1999}, as previously described in this section by different models.  Despite the fact that NGC\,4330 
is located at $\sim$0.4 virial radius \citep[e.g.][]{chung-2007}, its mass distribution is probably not affected by the cluster 
environment \citep{amram-1992} and can be considered as unperturbed.  Following \citet{lequeux-1983}, the actual dynamical mass 
of a spiral galaxy that contains a stellar disc, a spherical stellar bulge, and a spherical dark matter halo ranges in the mass 
interval given by $M(R)=\alpha RV_{rot}^{2}(R)/G$, where $G$ is the universal gravity constant,  $V_{rot}$ the flat rotation 
velocity, and $\alpha$ a parameter spanning between the lowest value of $\alpha$=0.6 for a galaxy dominated by a flat disc 
component and the higher value of $\alpha$=1.0 if the spherical halo components dominate. Inside the radius $R_{23.5}(i\mathrm{-band})\sim$11.6\,kpc, using $V_{rot}^{max}\mathrm{=158}\pm \mathrm{12\, km\,s}^{-1}$ and using $\alpha$=0.8$\pm$0.2 to be conservative, the dynamical mass of 
NGC\,4330 within $R_{23.5}(i\mathrm{-band})$ is $M_{dyn}(R_{23.5,i-band})$=4.4$\pm$\,0.6$\times$10$^{10}$ $\mathrm{M_{\odot}}$.  For the same maximum 
rotation velocity, the mass comprises within the HI radius $R_{HI}$=11.3\,kpc is  $M_{dyn}(R_{HI})$=4.9$\pm$1.2$\times$10$^{10}$ $\mathrm{M_{\odot}}$.
These estimates can be compared to the stellar mass of the galaxy, $M_{star}$\,=\,5$\times$10$^{9}\mathrm{M_{\odot}}$ (see Table \ref{general}). 

\subsection{Comparison with HI and CO kinematical data}

\begin{figure}[ht!]
\centering
\includegraphics[width=1\linewidth]{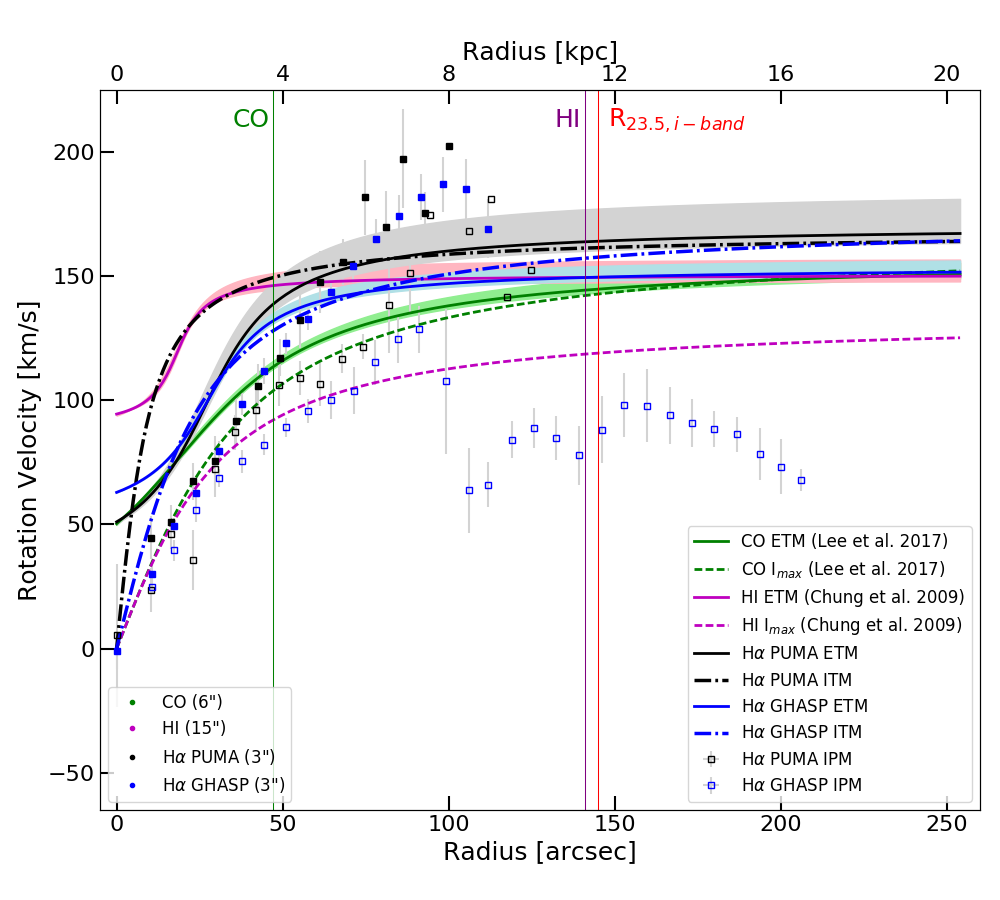}
\caption{
Rotation curves of NGC\,4330 derived using
the $^{12}$CO($2-1$) at $\lambda$=1.3 mm data from 
\citet{lee-2017} (green), the HI ($\lambda=$21 cm) data from 
\citet{chung-2009} (purple), and the FP H$\alpha$ data from PUMA (black) and GHASP (blue). The RCs from CO and HI data were computed using the envelope-tracing method (ETM).  
The solid lines correspond to the best fit to the points of the terminal velocity using $\eta$=0.3 and the shaded areas show the rotation velocity amplitude for $\eta=[0.2, 0.5]$ 
(see relation \ref{sofue-ec-intensity}); the dashed lines show the RC computed from the PVD maximum intensity.
For comparison, the H$\alpha$ RCs shown in Fig. \ref{RCET} are also reported on this plot. The green, purple and red vertical lines indicate the CO extension ($\sim$47 arcsec, $\sim$3.8\,kpc), the HI radius ($\sim$141 arcsec, $\sim$11.3\,kpc) and the outermost photometric radius ($R_{23.5}\sim$ 145 arcsec, $\sim$11.6\,kpc), respectively.
The top-right insert indicates the angular resolution of each dataset.}
\label{RC_CO_HI_Ha}
\end{figure}

In Fig. \ref{HIPVD} we show the HI velocity fields from \cite{chung-2009} and the RCs we derive from the PVDs. As expected, due to beam smearing effects, 
the RCs obtained using the minimum and maximum HI surface brightnesses strongly differ. These data are used to compute the ETM RC, which matches fairly well with 
the H$\alpha$ RC, except maybe within the first two kpc where the HI velocities have probably been overestimated (see Fig. \ref{RC_CO_HI_Ha}). 
The comparison between the H$\alpha$, HI and CO velocity fields are shown in Figs. \ref{hiha} and \ref{COHA} in the appendix, respectively.  The HI velocity field is the most 
extended with a diameter of $\sim$282\,arcsec ($\sim$22.6\,kpc) from the northeastern top to the southwestern lowest corner. The H$\alpha$ velocity field has been measured on a 
diameter of $\sim$205\,arcsec ($\sim$16.4\,kpc), and exceeds by more than a factor of two  the one derived using CO data ($\sim$94 arcsec, $\sim$7.5\,kpc). Despite these differences,
which might be partly due to the different sensitivity in the data combined with their different angular resolution (15\,arcsec for HI and 6\,arcsec for CO), the three velocity fields are comparable indicating a similar kinematics of the three components.
We applied the ETM to the $^{12}$CO ($2-1$) data of \cite{lee-2017} and to the HI data of \citet[][see Sect.  \ref{Subsect.Multifreq}]{chung-2009}.  
From the PVD extracted with a pseudo-slit along the kinematic major-axis we computed the terminal velocity using $\eta$=0.3, where  $\eta$ is a fraction of maximum intensity to set the envelope intensity  (see Eq. \ref{sofue-ec-intensity}), then we fitted 
the data to the  arc-tangent model (Eq. \ref{eq-vr-teorica}) using a least-squares minimization. The HI, CO, and H$\alpha$ ETM RCs are plotted in Fig. \ref{RC_CO_HI_Ha},
which also includes the H$\alpha$ RCs computed with the ITM and the IPM.

For the HI RC, the ETM fit gave a coefficient of determination $R^{2}=0.97$, with $r_{0}=18\pm 2$\,arcsecond $r_{t}=6\pm2$ arcsec, providing a RC with the steepest inner slope, represented 
by the purple line in Fig. \ref{RC_CO_HI_Ha}. This cuspy inner slope is due to the low HI surface brightness, which is reached thanks to the low-resolution VLA 
HI data ($\sim25$ arcsec, see Sect.\,\ref{Subsect.Multifreq}). This low-resolution induces strong beam smearing effects and an overcorrection at the lowest HI levels.  
The ETM provides also a way to estimate of the lowest limit of the RC using only the maximum intensity $I_{max}$ (in setting $\eta=I_{min}=0$, see Eq. \ref{sofue-ec-intensity}). 
This curve is represented by the purple dashed line in Fig. \ref{RC_CO_HI_Ha}. The interval between the two purple lines delimits all the possible HI RCs that could be computed 
and gives inner slopes compatible with those obtained at other wavelengths. The maximum HI rotation velocity is $\sim$10  $\mathrm{km\ s^{-1}}$ lower than those obtained with 
H$\alpha$ data. 

Since the CO is detected only in the inner regions, we have extrapolated its ETM rotation curve, using the parameters of the arctan function fitted to the observational data (see Sect. \ref{The envelope-tracing method}), to compare it with those computed at other wavelengths using the same method. The CO RC is represented by the green dashed line in Fig. \ref{RC_CO_HI_Ha}. The fit of the CO RC, which was obtained with a coefficient of determination $R^{2}=0.76$, gave $r_{0}=19\pm 10$ arcsec, and the transition between the rising and flat part of the RC at $r_{t}=33\pm26$ arcsec. These values are in agreement with those determined for the H$\alpha$ RCs using the same method. At large radii, the CO and the HI RCs match and have values $\sim$10 $\mathrm{km\ s^{-1}}$ lower than those derived using the H$\alpha$ data. We recall, however, that due to its limited extension, we do not have any observational constrain on the CO RC outside $\sim$50\,arcsec ($\sim$0.4\,kpc).

\section{Kinematical analysis from 2D-maps}
\label{section5}

\begin{figure*}[ht!]
\centering
\includegraphics[width=0.49\linewidth]{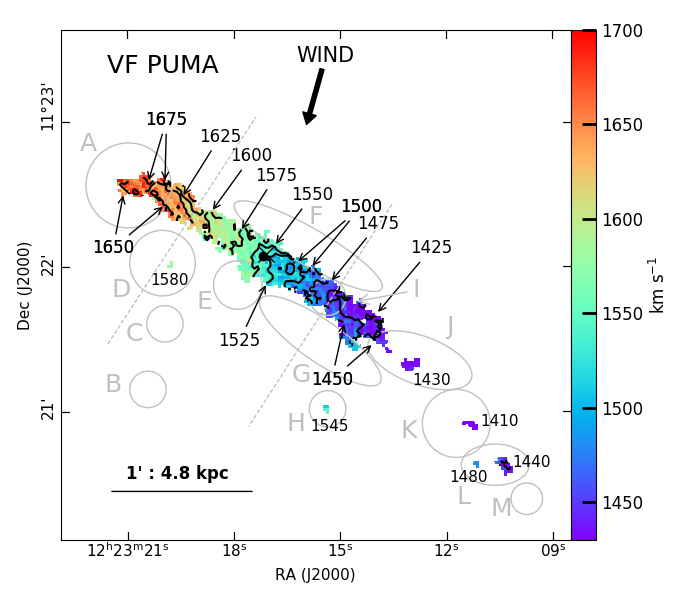}
\includegraphics[width=0.49\linewidth]{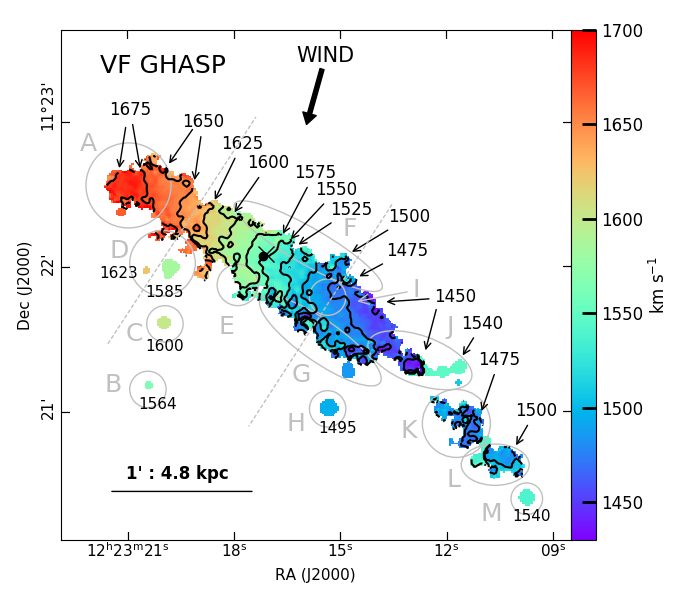}
\caption{Velocity fields of the ionised gas of NGC\,4330 (left PUMA, right GHASP). The general features of the main disc regions (northeast, central and southwest), identified by two parallel dashed lines, and the regions with particular features, identified with letters A to H, are discussed in the text. 
The black bullet and the cross indicate the position of the photometric and kinematical centre, respectively. The black arrow shows the wind direction as derived from the models 
of \citet{vollmer-2012-a, vollmer-2021}. 
}
\label{velocityfields}
\end{figure*}

\subsection{Velocity fields}

Figure \ref{velocityfields} shows the velocity field of the ionised gas of NGC\,4330 derived from the PUMA and GHASP FP datasets. The velocity fields of  both  datasets are  consistent, they display a rigid body shape inside of a diameter of $\sim$60\,arcsec ($\sim$4.8\,kpc) where the isovelocity contours are parallel to the minor axis of the galaxy. We have traced on these maps two parallel dashed lines at at $r_{0}\pm$ 30 arcsec ($\pm$2.4 kpc)  enclosing this region.  Outside this diameter the isovelocity contours become almost parallel to the major axis in the NW hook (region A) and on its counterpart, between regions I and J. However, these maps differ in the southern tail, i.e. in regions K and L (see point 5 hereafter). These last features are the signature of non-circular motions that indicate strong perturbations in the kinematics of the regions affected by the stripping process.
The two-dimensional velocity field obtained with high-resolution FP interferometry allows us also to measure the kinematic of the gas in very low surface brightness regions. These regions, labelled from A to M, were previously identified in the monochromatic image and they are now identified on the velocity fields of both datasets 
in Fig. \ref{velocityfields}. Our data allowed us to measure the LoS velocities of the stripped gas in the southern regions, in region F located on the compression front on the northeaster side of the disc and in region I on the disc plane. The LoS velocities of all these regions are given in Table \ref{TableKinematicalProperties} and will be further discussed in Sect. \ref{PeculiarMotions}.

We notice that:
\begin{enumerate}
\item  The detection and estimation of the LoS velocity of the gas over the galaxy disc, and more specifically inside of regions A, E, F, G, H, I, J (westernmost part) are robust since derived from independent sets of data with consistent results.

\item  The gas inside region A has a radial velocity of $\sim$1650 km~s$^{-1}$ that is $\sim$\,25\,km\,s$^{-1}$ lower than that of the disc at the same galactocentric distance.

\item  If we compare the radial velocities in the southeastern regions of the galaxy with respect to the radial velocities in the plane of the disc, either with respect to their position perpendicular to the plane of the disc, or with respect to the direction of the wind as deduced from the direction of the tails and predicted by simulations \citep[e.g.][]{vollmer-2021}, we notice that the gas inside regions B, C, D, and H has radial velocities significantly different ($\sim\pm$50\,km\,s$^{-1}$) from those measured in the disc from where the gas has been stripped. The radial velocities of the gas regions B, D, and C is lower than that in the disc, while the radial velocity measured in region H is higher by the same amount.

\item Despite the fact that they are above the FP detection limit observed in other regions, the detection of ionised gas in regions B, C, D, M and in the western tail of region J  could have been affected by night skyline contamination. Indeed, the night-line subtraction was very difficult because the OH-radical has a strong emission line at $\sim$6597\,\AA\ 
\citep[e.g.][]{Osterbrock-1996PASP..108..277O},
roughly at the same wavelength that the redshifted H$\alpha$ emission. On the other hand, 
the LoS velocity of the gas in region H, which has been probably originated in the approaching side of the disc, is lower and different from the LoS velocity of the gas 
measured in regions B, C, and D, in the receding side. This difference suggests that these estimates are not largely affected by the residual night skyline emission.

\item  
Ionised gas is detected in regions K and L with similar LoS velocities. A velocity difference of $\sim$50\,km\,s$^{-1}$ is observed between both datasets, which could also be due to parasitic night skylines. 
Thus, despite this difference in the two datasets, these velocities are consistent with streaming motions caused by the stripping process.

\item The  gas detected in regions J, K, and H  has a velocity similar to that measured at the edge of the rotating disc (1450\,km\,s$^{-1}$), suggesting a flat rotation curve.

\item Figure \ref{velocityfields} also shows two dashed lines at $\sim$30\,arcsec ($\sim$2.4\,kpc) from the galactic centre. 
These dashed lines enclose the disc section where the ionised gas shows a rigid-body rotation as indicated 
by the velocity contours parallel to the minor axis. To the southwest, region I marks the limit where the velocity contours start to become parallel to the major axis. Its 
counterpart to the northeast is the beginning of the  hook shape where the isovelocities become not parallel to the minor axis. Region I might represent the outer layer of the vertical disc stratification, at the interface with the surrounding medium, where ram pressure begins its action. Since the stripping 
pressure is not perfectly vertical on the disc, the I region suffers a radial component of the stripping force.

\end{enumerate}

In order to check the highly uncertain velocity of region D gathered with the FP data, we gathered a long-slit spectrum on February the 17, 2021 using the low-resolution ($R$=700) MISTRAL spectrograph installed on 
the 193 cm OHP \citep{adami-2018}. We took four independent exposures of 15 minutes each, for a total of 60 minutes. The exposures were gathered using a slit 
of width 1.9\,arcsecoriented North-South to include at the same time region D and the disc of the galaxy. The low-resolution spectrum of region D is shown in 
Fig. \ref{mistral}.  
Because of the night skyline contamination, the H$\alpha$ line is marginally detected with a signal-to-noise ratio $\sim$2 and a mean LoS velocity of 
1458$\pm$92\,km\,s$^{-1}$, only marginally consistent with the FP measurements of 1580$\pm$30\,km\,s$^{-1}$.
 Figure \ref{mistral} also shows the corresponding MISTRAL spectrum obtained on the galaxy disc north of region D, where the average velocity is 1629$\pm$46\,km\,s$^{-1}$, 
 in very good agreement with the FP data. As for the other extraplanar regions detected in the FP data, the velocity in the stripped gas is lower than that 
 measured in their corresponding regions to the north on the galaxy disc.

The gas stripping process removes the gas from the disc and transfers and dilutes it into the intracluster medium. 
The only velocity component that we can measure is the one along the LoS.  For edge-on galaxies it is straightforward to 
interpret the LoS velocity as rotation velocity in the galaxy disc plane. The intergalactic wind might change the game, in adding a velocity component related to the direction 
of motion of the galaxy within the cluster.  Nevertheless, in the specific case of NGC\,4330, the wind is almost perpendicular to the galaxy disc and has a large component in 
the sky plane, as suggested by the numerical simulations (see Sect. \ref{section7}). For these reasons the observed velocities of the gas expelled from the disc are not expected to 
dramatically change. Thus, the gas continues to rotate outside of the disc and shows the same LoS velocities than those
it had before being stripped. Since regions B, C, D (H) have LoS velocities lower (higher) than their corresponding region in the disc, this suggests that here the 
gas partly lost its angular momentum during the ram pressure stripping event.

\begin{figure*}[ht!]
\includegraphics[width=0.5\linewidth]{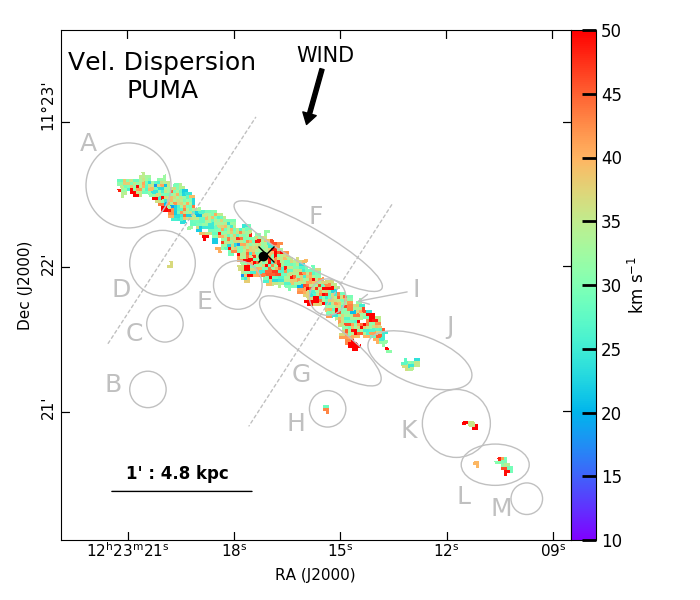}
\includegraphics[width=0.5\linewidth]{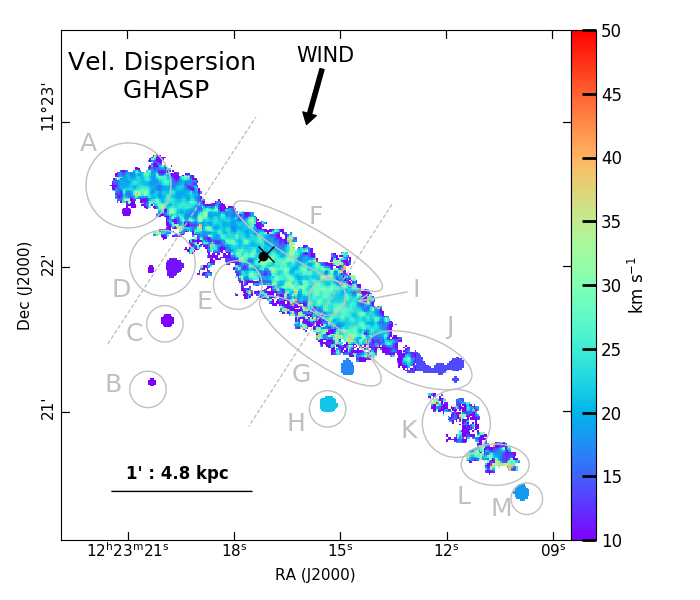}
\caption{Velocity dispersion fields of NGC\,4330 (left PUMA, right GHASP). The general features of the main disc regions (northeast, central and southwest), identified by two parallel dashed lines, and the regions with particular features, identified with letters A to H, are discussed in the text. The black bullet indicates the position of the photometric centre. The black arrow shows the wind direction as derived from the models of \citet{vollmer-2012-a, vollmer-2021}.}
\label{dispersion}
\end{figure*}

\begin{figure*}[ht!]
\includegraphics[width=0.5\linewidth]{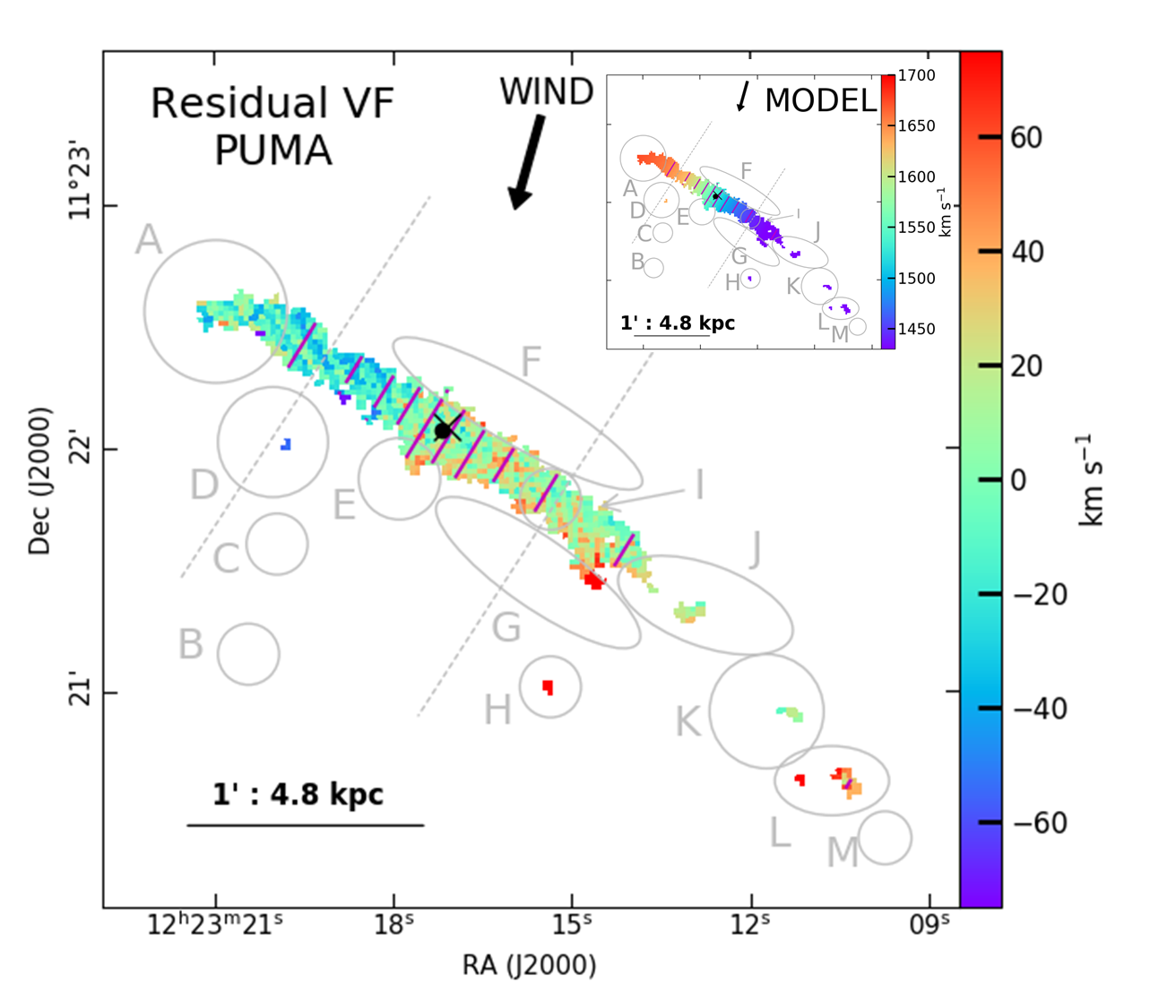}
\includegraphics[width=0.5\linewidth]{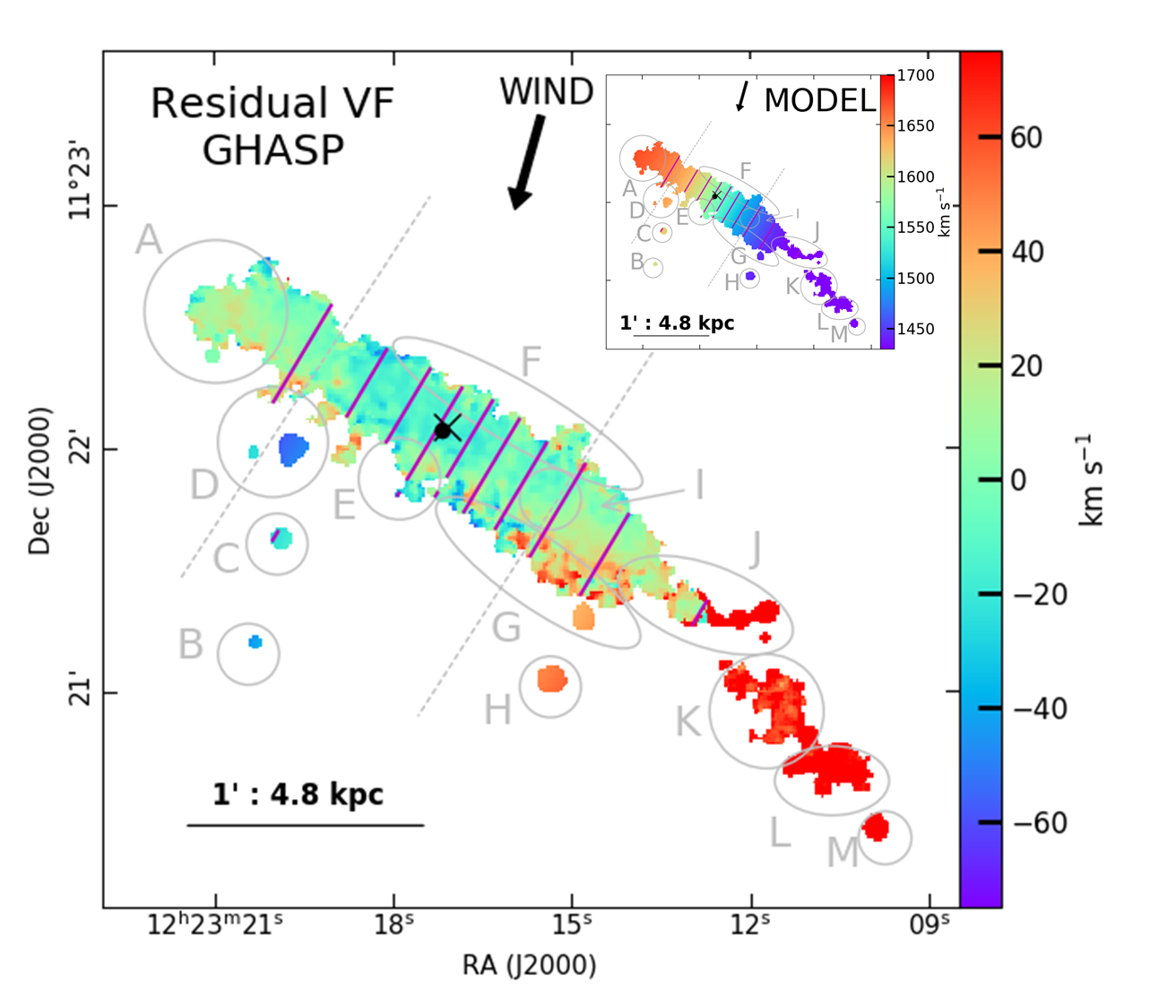}
\caption{Residual velocity fields obtained after subtracting the velocity field computed from the ITM model data cube (left PUMA, right GHASP). 
The model, based in a solid-body rotation, is shown in the inset. The pink lines show the isovelocity contours of the edge-on rotating disc model. The bullet and the cross indicate the position of the photometric and kinematical centre, 
respectively. The arrow indicates the wind direction as derived from the models of \citet{vollmer-2012-a, vollmer-2021}.}
   \label{residual}
\end{figure*}

\subsection{Velocity dispersion}

Figure \ref{dispersion} shows the velocity dispersion of the galaxy as measured from the PUMA and GHASP datasets.
Following \cite{boselli-2021} we checked that the velocity profiles have a single peak, i.e. that the barycentric method applied to measure  the velocity dispersions does not 
introduce any strong systematic bias in the results.
  
At first order, the velocity dispersion of the gas is fairly constant over the disc of the galaxy, with $\sigma$ $\sim$\,35$\pm$10\,km\,s$^{-1}$ and $\sigma$ $\sim$\,21$\pm$7 
km s$^{-1}$ for the PUMA and the GHASP data, respectively.
The PUMA velocity dispersion is higher than the one measured in the GHASP data because of its reduced spectral resolution.  It is indeed difficult to measure velocities 
far below the instrumental resolution ($\sigma_{LSF}$\,=\,19.5 and 13.1\,km\,s$^{-1}$, respectively) as it has been recently discussed in \cite{boselli-2021} in comparing 
VLT/MUSE data with FP data obtained with GHASP.

At higher spatial frequencies, fluctuations in the velocity dispersion are observed on scales comparable to or larger than the one set by the seeing.  
These fluctuations are smoother in the GHASP data than in the PUMA ones. This is due to the difference in the method used to compute the map (Gaussian smoothing versus 
Voronoi tessellation, respectively).

Panels (c) and (e) of Fig. \ref{henri} show intensity-velocity dispersion diagnostic diagrams and panels (a) and (d) the histograms for the intensities 
and of the velocity dispersions. These diagnostic diagrams are generally used to characterise the dynamics of the ISM in extragalactic giant HII regions 
\citep{munoz-tunon-1996} and also to investigate the dynamic of HII galaxies \citep{moiseev-2012, carvalho-2018}. \citet{moiseev-2012} show 
a sketch where different features and regions in the plot can be interpreted in terms of (case 1) HII regions (constant velocity dispersion 
region with high H$\alpha$ intensity), (case 2) shell (inclined bands in the plot), (case 3) low density turbulent ISM (triangular shape  
with a base spanning from low to high velocity dispersion and a triangle height ranging from low to moderate H$\alpha$ intensity). 

Shells (case 2) are not observed in the diagnostic diagrams shown in panels (c) and (e), probably because of the lack of spatial resolution for this 
relatively distant galaxy. The triangular shape distribution of the diagnostic diagram (case 3) is asymmetric, with a deviation toward the lowest 
velocity dispersions at low intensity and toward larger velocity dispersions at intermediate intensities, indicating the supersonic behaviour of 
the ISM in NGC\,4330. In panel (e) we selected four areas, represented by ellipses, of different pixel density (from high to low mean pixel density: red, green, 
dark, and light blue), and identified their locations in the galaxy in panel (b). The red pixels with subsonic velocities are mainly located on the 
outskirts of the galaxy while the dark blue ones, with supersonic velocities, are a little bit on the inner side (case 3). The supersonic green pixels 
are located along the middle plane of the galaxy while the light blue with the highest intensities (case 1), match with the nucleus of the 
galaxy and the brightest HII regions. Their velocity dispersion is between 20\,km\,s$^{-1}$ and 26\,km\,s$^{-1}$. 
 
\subsection{The residual velocity fields}

Two-dimensional (2D) residual velocity fields are generally used to optimise the free parameter determination of RCs that are computed from a 2D velocity fields. In general, 
because the RCs are based on axisymmetric models, a pattern observed on the residual velocity field would be due to incorrect parameter determination. As a consequence, when 
the free parameters of the rotation curve are optimised, the residual velocity field is pattern-free and the dispersion in the 2D residual velocity is minimum 
\citep[see][]{warner-1973, fuentesc-2004, erroz-2012}.

The PUMA and GHASP residual velocity maps ($V_{res}=V_{\rm{LoS}}-V_{mod}$) shown in Fig. \ref{residual} were obtained after subtracting the iteration model velocity 
field ($V_{mod}$) from the observed one ($V_{\rm{LoS}}$).  Once we identified the regions presenting a large scatter with the mean velocity dispersion of the residual velocity 
field, which are not supposed to be in rotation in the plane of the disc, they were masked from the observed velocity field and the axisymmetric model was recomputed. This is because 
the axisymmetric model is not supposed to be observationally constrained by non-axisymmetric motions.

The residual velocity maps given in Fig. \ref{residual} do not show any systematic pattern in the galactic disc and have a velocity dispersion of 
$\sigma_{res}\sim 19\, \mathrm{km\,s^{-1}}$ for the GHASP dataset and $\sigma_{res}\sim$\,24\,km\,s$^{-1}$ for the PUMA dataset. 
The residual velocity is $\sim0\,\mathrm{km\,s^{-1}}$, suggesting that the adopted cylindrical model (described in section \ref{ITM}) reproduces well the axisymmetric rotation of the disc modelled by 
the rotation curve and its associated parameters. 

An accurate examination of Fig. \ref{residual} indicates how the kinematical properties of the various extraplanar ionised gas features created after a ram pressure stripping 
event differ from  those properties predicted by a purely axisymmetric cylindrical model. The most discrepant regions in terms of relative velocity of the gas are regions G, 
H, and L for the PUMA data and G, H, and from J to M on the southwest side of the galaxy, for the GHASP data. Here the mean gas velocity is $\sim$50\,km\,s$^{-1}$ higher than that observed 
in the model.
A less pronounced difference in the same direction is observed in region E ($\sim$\,20\,km\,s$^{-1}$), which is located in projection close to the kinematic centre of the disc. 
On the contrary, the gas inside the regions D (PUMA) and B, C, D (GHASP) has a mean velocity $\sim$\,40\,km\,s$^{-1}$ higher than that of the cylindrical model. 
Additional analysis about these individual regions is given in Sect. \ref{PeculiarMotions}.

\subsection{Peculiar motions}
\label{PeculiarMotions}
%

\begin{table}[ht!]
\begin{center}
\caption{Velocities of peculiar regions derived from the PUMA and GHASP datasets.}
\begin{tabular}{c|ccc|ccc}
\hline\hline
& \multicolumn{ 3}{c|}{PUMA} & \multicolumn{ 3}{c}{GHASP} \\ 
 \hline
Region  & $V_{rad}^{(1)}$ & $\langle\sigma\rangle^{(2)}$ & $V_{res}^{(3)}$ & $V_{rad}^{(1)}$ &  $\langle\sigma\rangle^{(2)}$ & $V_{res}^{(3)}$ \\ 

 \hline
A & 1659 & 35 & -8  & 1673 & 19 & 8 \\ 
B & ... & ... & ... &  1564 & 9 & -41 \\ 
C & ... & ... & ... &  1601 & 9 & -21 \\ 
D  & 1580 & 37 & -54 &  1585 & 11 & 34 \\ 
E  & 1562 & 38 & 20 &  1579 & 19 & 2 \\ 
F  & 1543 & 38 & 21 &  1533 & 20 & -2 \\ 
G  & 1477 & 53 & 48 & 1493 & 19 & 22 \\ 
H  & 1548 & 37 & 109 &  1496 & 21 & 51 \\ 
I  & 1460 & 37 & 13 &  1482 & 25 & 3 \\ 
J  & 1435 & 33 & 19 &  1542 & 14 & 116 \\ 
K  & 1412 & 45 & 9 &  1480 & 15 & 72 \\ 
L  & 1450 & 38 & 49 &  1501 & 21 & 99 \\ 
M  & ... & ... & ... &  1541 & 18 & 139 \\ 
\hline
\end{tabular}
\end{center}
\tablefoot{
$^{(1)}$ Mean line-of-sight velocity in km s$^{-1}$. 
$^{(2)}$ Mean velocity dispersion in km s$^{-1}$. 
$^{(3)}$ Mean residual velocity in km s$^{-1}$. 
}
\label{TableKinematicalProperties}
\end{table}

Table \ref{TableKinematicalProperties} gives the main kinematical properties of the gas in regions A to M. The GHASP data indicate that the gas inside most of these particular regions has a velocity dispersion near to the mean value measured over the galactic disc, and this 
is probably due to the Gaussian smoothing used to create the map of the velocity dispersion. The PUMA data rather indicate that the gas inside all the same regions has a 
velocity dispersion higher than the one measured within the galaxy disc, with the higher velocity dispersions measured in regions E, G, and K. On the other hand, the residual 
velocity field derived from both sets of data shows that the gas in regions A, E, F, and I follows the rigid-body rotation with a low residual velocity. This result is 
expected for the gas in regions E, F, and I because they are still associated to the galactic disc, but not for the region A which has been perturbed by the stripping 
process.

In region A, the velocity contours from both datasets indicate that the gas is rotating with a slight bending with respect to the minor axis. The PVD obtained with the PUMA 
dataset shows that the gas at the northeast, where the H$\alpha$ monochromatic map has a hook feature, has higher radial velocities than those expected from the model. 
For this reason, we could expect that here the gas has a higher velocity dispersion due to non-circular motions, corresponding to high values on the residual velocity map. 
However, the RCs computed with IPM, which show the gas rotating as a rigid body in region A, do not confirm this result. A decrease of the velocity rotation with respect 
to the rigid-body rotation is instead observed in the RC and in the residual velocity maps in the southwestern side of the galaxy.

Of particular interest is region I, where the GHASP data indicate that the velocity dispersion is  $\sim5\ \mathrm{km\ s^{-1}}$ higher than the mean value estimated over 
the disc of the galaxy. The gas inside region I still follows the rigid-body rotation as indicated by a low residual velocity in both maps. Region I is located at the boundary 
between the rising and the flat part of the RCs, where the velocity contours in the velocity fields starts to become parallel to the major axis. The PVD shows that here the 
ionised gas emission start to decrease significantly, and that outside this region the gas starts to rotate slower than expected by the model. Its counterpart on the 
northeast side of the disc also shows a drop in surface brightness in the PVDs, but outside this region the rotation velocity of the gas seems to increase with respect to the model. 
In this symmetric region at the northeast of the disc, the gas has no high velocity dispersion and no high residual velocity values. Therefore, the increase of the line width 
of the gas observed in region I might be due to its acceleration, which is able to remove it from the galaxy disc in the southwest of the galaxy during the stripping event.

Figure \ref{velocityfields} shows that the gas inside regions B, C, and D has a lower LoS velocity with respect to those measured in their corresponding positions on the disc plane 
while region H has a higher one. This means that these four regions rotate slower than their counterparts on the galactic disc. This can also be seen on the residual velocity 
field  (Fig.e \ref{residual}) where the velocities are given in the rest frame of the rotating disc and in which negative (positive) velocities on the approaching (receding) 
side indicate a lower amplitude rotation outside the plane of the disc. The gas in regions B, C, D, and H does not have a high velocity dispersion, 
meaning that the wind does not bring a substantial quantity of kinetic energy to those regions. This probably indicates that the gas of these regions had lost angular 
momentum when it was detached from the galactic disc.

At the southern edge of the galactic disc, the velocity contours are bent with respect to the minor axis. This is particularly evident for the gas associated to region G, 
where the velocity dispersion and residual velocity increase smoothly with galactocentric distance, reaching their largest value in the detached cloud 
at the southwester edge of the region.  The same cloud at the edge of region G has also a higher LoS velocity than its counterpart on the disc, which means a lower rotation 
velocity because this cloud is located on the receding side of the disc.  Regions B, C, D and H show a similar behaviour, they display lower LoS velocities, thus lower rotation 
velocities because they are approaching to the observer with respect to the galaxy centre. Then, we can infer that the gas inside regions J to M, with a low velocity dispersion 
but a high residual velocity, has evolved in a similar way than the gas in regions B, C, D and H, i.e. having lost angular moment once stripped from the disc. Finally, the gas in region L has a higher velocity dispersion that the one observed around regions K and M. This does not mean that it gets kinetics energy from the wind because 
the line broadening is probably due to the presence of a star forming region (see Fig. \ref{Hacomposite}).


\section{Kinematics of the gas out the plane}\label{section6}

\begin{figure*}[ht!]
\centering
\includegraphics[width=1.\linewidth]{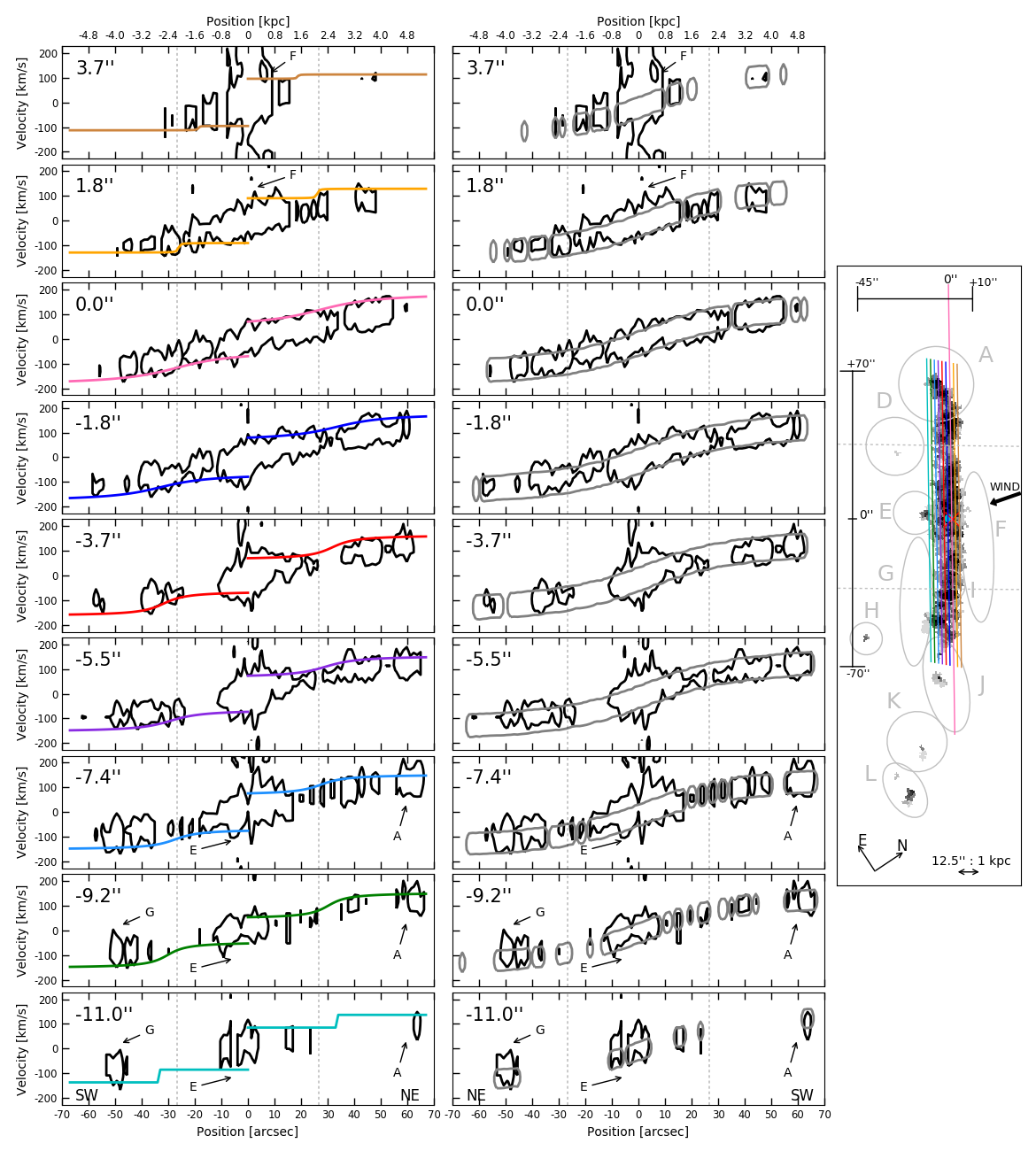}
\caption{Position-velocity diagrams (PVDs) parallel to the kinematic major axis extracted from the PUMA dataset. The distance of the pseudo-slit form the disc plane 
along the $z-$axis (positive to the north) is given in the upper-left corner of each panel. The positions of all the pseudo-slits are shown on the H$\alpha$ monochromatic 
map in the right panel.  From the top at +3.7  to -5.5\,arcsec (from $\sim$0.3 to $\sim$0.4\,kpc) the intensity of the contour level is at 1.5 $\sigma$, from -7.4 to -11 
arcsec (from $\sim$0.6 to $\sim$0.9\,kpc) the intensity level is at 1.2 $\sigma$. In the left panels, coloured lines overlaid on the PVD represent the best fit of the 
terminal velocity (relation \ref{sofue-ec-intensity} with $\eta=0.3$) using the arc-tangent model (Eq. \ref{eq-vr-teorica}). There is a correspondence in colour 
between the terminal velocities and the pseudo-slits plotted on the monochromatic map in the right panel. In the middle panels, the PVDs derived from the data cube 
model, that was built in the context of the iteration method to compute the rotation curve (see Sect. \ref{ITM}), are overlaid on the observed PVD in gray. All the 
PVDs extracted from the data cube model are at 8\% of the maximum intensity of the model. 
The two parallel dashed lines are at $r_{0} = \pm$ 30 arcsec ($\pm$ 2.4 kpc) and indicate the three main disc regions: northeast, central and southwest.  The different features of interest are indicated with letters A to H.
}
\label{Figure-PVDs}
\end{figure*}

In this section we focus on the PVDs at different locations in the galaxy disc, parallel and perpendicular to the major axis of the galaxy, in order to 
study the kinematics of 
the gas out of the plane of the stellar disc. Figures \ref{Figure-PVDs} and \ref{pvd-minorA}, on which we have also located regions A to M, show the PVDs extracted from the PUMA data cube measured along the major and the minor axis respectively, at different distances from disc plane 
\citep[e.g.][]{epinat-2008-b, Rosado-2013}. 

\subsection{PVDs along the major axis}

Figure \ref{Figure-PVDs} shows the PVDs along the major axis at different distances from the disc plane. The pseudo-slits are placed along the z-axis from +3.7\,arcsec
($\sim$0.3\,kpc) to the northern direction (up wind direction) of the kinematic major axis, up to -11\,arcsec ($\sim$0.9\,kpc) to the southern direction (tail direction), 
and they are separated each other by $\sim$1.8\,arcsec($\sim$0.1\,kpc, see Sect. \ref{Sec:pvds}). Figure \ref{Figure-PVDs} shows that the rotating gas is less extended along the disc upwind (+1.8 arcsec), 
where it reaches
50\,arcsec ($\sim$4\,kpc) in both the northeast and southwest directions, than in the tail (70 arcsec to the northeast and 60 arcsec to the southwest). This result is expected 
in a ram pressure stripping scenario, where the gas is removed outside-in 
\citep[e.g][]{Quilis-2000Sci...288.1617Q, vollmer-2001, boselli-2006-b, boselli-2021, Roediger-2005A&A...433..875R, Tonnesen-2009ApJ...694..789T}. 

To identify the main kinematic features resulting from the ram pressure stripping event, we compare the predicted values by theoretical fittings with the observed data. We have applied the 
ETM on each PVD measured along the major axis at different distances from the disc plane. In the left panels of Fig. \ref{Figure-PVDs} the best fit to the terminal velocity of each PVD is overlaid on its correspondent plot. In addition, in the context of the 
iteration method (ITM),  we built a first order approximation data cube model to compute the rotation curve (RC) of NGC\,4330 (see in Sect. \ref{section4}). Here we compare the 
PVDs from the observational data with the PVDs at the same position obtained from that data cube model in the central  panels of Fig. \ref{Figure-PVDs}. We notice that the 
cylindrical model follows the emission intensity of the galaxy in velocity and matches with the kinematical behaviour of the major axis PV diagram.

Some low surface brightness plumes of H$\alpha$ emission were detected on top of the dust lane inside region F. These plumes, which can be seen in the upper two panels of 
Fig. \ref{Figure-PVDs}, show an increment in velocity by $\sim30\ \mathrm{km\ s^{-1}}$ breaking the symmetry of the PVDs. If the H$\alpha$ emission, after being compressed 
on the front side of the galaxy during the stripping event, is pulled as a whole to the southwest of the stellar disc, these plumes represent gas dragged that is still resisting 
to change, giving us a clue of the initial state of the disc.

The PVD along the major axis presents a kinematic symmetry with respect to the kinematic centre without any plume or knot with important variations in velocity with respect 
to the model. The kinematic major axis matches the bent dust lane in the central region, within a radius of $\sim$30\,arcsec ($\sim$2.4\,kpc), enclosed by the vertical dashed lines. 
As it is suggested by the bending of the velocity contours (see Fig. \ref{velocityfields}), the radial velocities tend to be constant $\sim150\ \mathrm{km\ s^{-1}}$ to the southwest, 
while to the northeast, the PVDs show velocities that change by $\sim \pm25\ \mathrm{km\ s^{-1}}$ perpendicular to the disc. To the northeast, the gas in region A has a radial velocity 
of $\sim200\ \mathrm{km\ s^{-1}}$.
As indicated by the residual velocity map (see Sect. \ref{section4}), although the gas inside region E has a peculiar shape, it rather follows the rotation of the galaxy. 
On the contrary,  the gas in region G has a lower velocity than in the disc.

\begin{figure}[ht!]
\centering
\includegraphics[width=1.\linewidth]{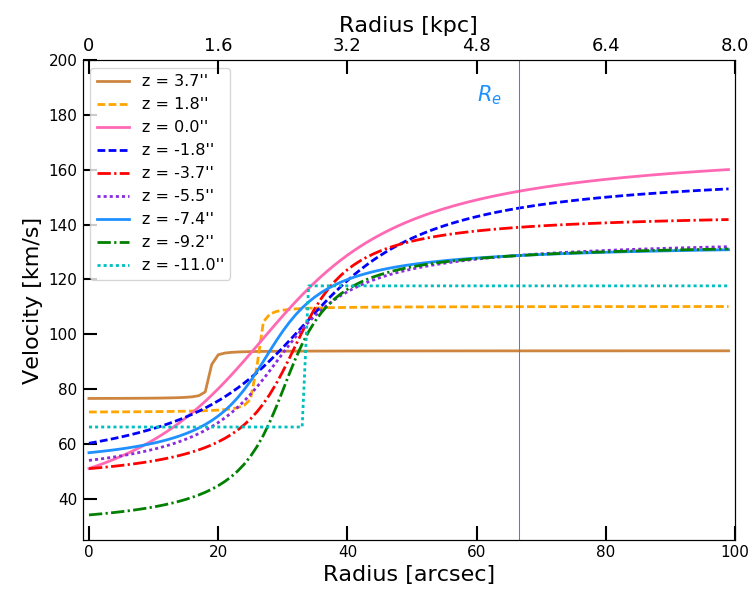}
    \caption{
Envelopes corrected from the morphogical inclination (89.5\degr) of the different PVDs shown in Fig. \ref{Figure-PVDs}. The vertical blue dotted line indicates the effective radius $R_{e}(i)$.
}
   \label{n4330_VrZ_521}
\end{figure}

\subsection{The Lag Gradient}

Kinematical studies of edge-on galaxies revealed the presence of a vertical velocity gradient in the rotation curve moving away from the disc plane.
This gradient has been observed at optical \citep[e.g.][]{heald-2006, Rosado-2013, bizayaev-2017, levy-2019} and radio \citep[e.g.][]{miller-2003, fraternali-2006, Zschaechner-2015-b} 
wavelengths, including in the Milky Way  \citep[e.g.][]{kalberla-2014}.  
This phenomenon has been explained by a combination of vertical movements due to energy input from galactic fountains, radial movements of the gas due to the pressure gradient of the 
halo and a declining rotation velocity due to the conservation of the angular momentum \citep{fraternali-2006, heald-2007}. This vertical velocity gradient can be modified whenever a galaxy such as NGC\,4330 is suffering an external perturbation. In NGC\,4330 the amplitude of the rotation curve does not decrease sufficiently with the vertical distance from the disc plane to explain the observed lags.  In this galaxy, which is suffering an almost face-on ram pressure stripping event, galactic fountains were probably suppressed, in particular in the upwind side of the disc where the external pressure is significantly increased.

The lag gradient of NGC\,4330 can be estimated by measuring the change in velocity along the $z-$axis using the PVDs parallel to the major axis.
Once corrected from the morphological inclination of the galaxy (89.5\degr), the envelopes of the different PVDs shown in Fig. \ref{Figure-PVDs} become the rotation curves given in Fig.  \ref{n4330_VrZ_521}.  
The rotation curve with the highest velocity amplitude is obtained in the plane of the disc ($z$=0 arcsec). To the southwest, at the effective radius, the velocity decreases by $\langle\Delta V\rangle\simeq 9\ \mathrm{km\ s^{-1}}$ each $\Delta z$=1.8\,arcsec ($\sim$0.14\,kpc) from the kinematic major axis up to $z=-5.5$\,arcsec ($\sim$0.4\,kpc), where, because the ram pressure stripping compresses the gas, there is no change in velocity 
along the $z$-axis up to -11\,arcsec ($\sim$0.9\,kpc), where the most diffuse gas is. To derive the velocity gradient along the $z$-axis of NGC\,4330 we used the fact that the asymptotic velocity $V_{0}$ of the relation (\ref{eq-vr-teorica}) is one of the free parameters of the fitting to the terminal velocity of each PVD, so that, from 
$z=0$ to $-5.5$\,arcsec ($\sim$0.4\,kpc) there is a lag gradient of $\Delta V/\Delta z\sim 64 \pm 9\mathrm{km\ s^{-1}\ kpc^{-1}}$. 
The rotation in the north along the $z$-axis has been modified by the wind, it is slower by $\Delta V=49\ \mathrm{km\ s^{-1}}$ with respect to the one measured at $z$=0 arcsec. The difference in velocity between the northern RCs is  
$\Delta V=16\ \mathrm{km\ s^{-1}}$. Although the rotation velocity decreases with the distance from the stellar disc, as indeed expected, it is not possible to 
determine a constant lag gradient in this direction.

\subsection{PVDs along the minor axis}

\begin{figure*}[ht!]
\centering
\includegraphics[width=1\linewidth]{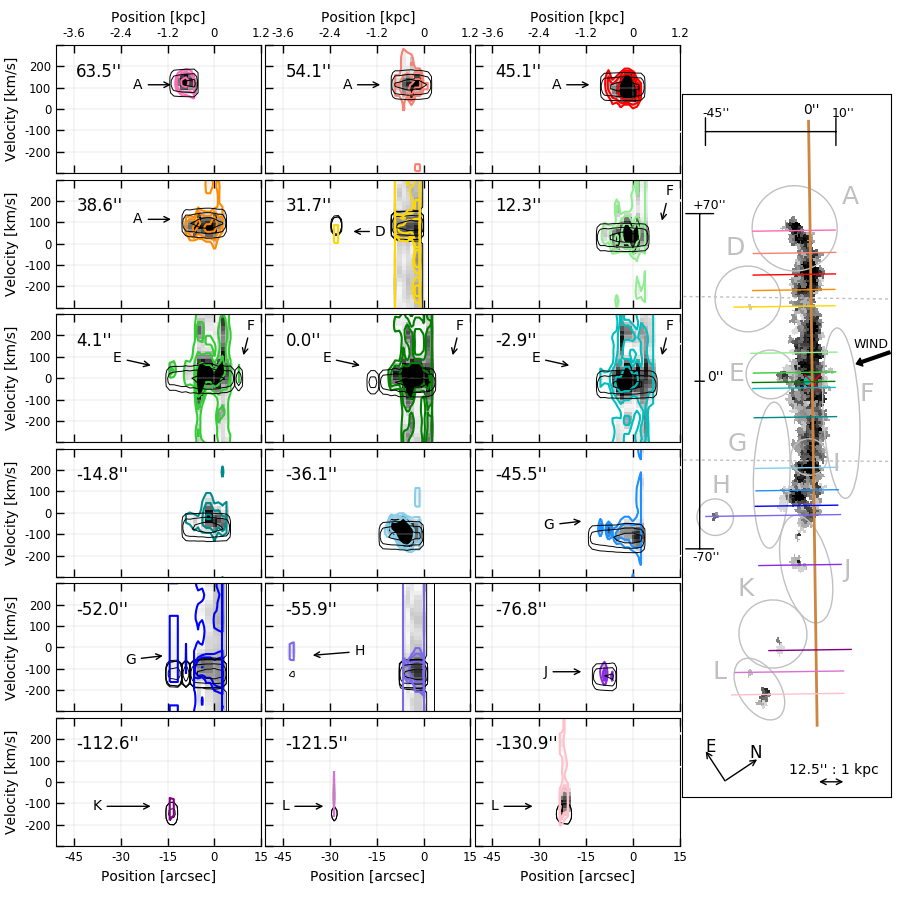}
\caption{Position-velocity diagrams parallel to the kinematic minor axis obtained from the PUMA dataset. The locations of the pseudo-slits are in the upper-left corner 
of each panel and they are traced on the H$\alpha$ monochromatic map in the left panel with colour correspondence. The distance increases from the 
south to the north of the disc, with zero at the kinematic centre. The black dotted ellipse on the H$\alpha$ monochromatic map indicates the $i$-band
$\mu_{e}(i)=23$ mag arcsec$^{2}$ isophote. The intensity of the contour levels is 1.3, 1.4 and  1.6 $\sigma$ from $-45.5$ 
to $+63$\,arcsec ($\sim -$3.6 to 5\,kpc) and 1, 1.1 and 1.25 $\sigma$ from $-130.9$ to $-52$\,arcsec ($\sim -$10.5 to $-$4.2 kpc). The PVDs extracted
from the data cube model (black contours) are at 1, 10 and 50\% of the maximum intensity of the model.
}
\label{pvd-minorA}
\end{figure*}

The ram pressure stripping event, which is acting almost face-on in NGC\,4330, has reduced the vertical height of the ionised gas disc in the northern direction (upwind).  
We thus study the PVDs along the kinematic minor axis to determine whether galactic fountains or inflows have been suppressed by the
external pressure \citep[e.g.][]{heald-2006, heald-2007, Rosado-2013}.
Figure \ref{pvd-minorA} shows the PVDs perpendicular to the kinematic major axis obtained from the PUMA dataset. 
The pseudo-slits are located in specific places to highlight the main features of the disc and its asymmetries, and they cross the regions labelled from A to M  
identified in Sect. \ref{section2}. 
We also compare  the PVDs from the observational data with the PVDs at the same position obtained from that data cube model computed with the iteration method 
(ITM, see Sect. \ref{section4}). 

In general, these PVDs show that the brightest H$\alpha$ emitting gas is compressed to the galactic mid plane and that it follows mostly the rotation predicted by the cylindrical model. 
Upwind (northeast) there is no H$\alpha$ emitting gas beyond $z\simeq 5$\,arcsec (0.4\,kpc).  
Only near the midplane of the disc, downwind (southwest) the gas decreases its brightness along the $z-$axis following the exponential structure used to construct the model. 
Here the velocity of the most diffuse gas generally increases with $z$ all over the galactic disc. This increase in velocity with $z$
is not reproduced by the ITM model.

In region A, at 65.5\,arcsec (5.2\,kpc) from the kinematic centre, the ionised gas has exactly the same velocity of the cylindrical model. At 31.7\,arcsec (2.5\,kpc), the low 
radial velocity observed for the gas in region D in the velocity field is now evident as long as its corresponding position on the disc follows the model. We only detected 
some diffuse gas to the northeast along the $z-$axis inside region F at a radius of 4.1\,arcsec (0.3\,kpc) from the kinematic centre, where the diffuse gas rotates a little faster than the model. The pseudo-slit at the kinematic centre shows 
that the gas in region E is completely compressed to the galactic mid plane. To the southwestern edge of the disc, the most diffuse gas inside regions G, H, J, and K rotates slower than what expected from the model, in agreement with the residual velocity map. 
Finally, we have observed that the gas inside region L (-121.5 and -130.9 arcsec, 9.7 and 10.5\,kpc respectively) has a high velocity dispersion, radial and residual velocity 
likely because the gas is moving with non-circular motions. Figure \ref{pvd-minorA} also shows that the diffuse gas in region L has a velocity higher by $\sim$\,$200\ \mathrm{km\ s^{-1}}$ than the one expected from the model.

Instead of a heart-shaped profile, which would indicate the presence of an inflow in the galaxy \citep[e.g.][]{fraternali-2006, Rosado-2013, Zschaechner-2015-a}, the PVDs suggest features similar to the intermediate- and high-velocity clouds of the Milky Way \citep[e.g.][]{waker-1997, fraternali-2006} whose origin is attributed to the interaction 
with the Magellanic Clouds. This suggests that the kinematics of the extraplanar diffuse ionised gas of NGC\,4330 is likely dominated by the ram-pressure stripping process.

\section{The dynamical model} \label{section7}

\label{The dynamical model}

We present here the kinematical results of the simulation of NGC\,4330 presented in \cite{vollmer-2021}.  They have been obtained using a N-body code, described in \cite{vollmer-2001}, which consists of a non-collisional component plus a collisional one.  The non-collisional component of 81\,920 particles simulates the stellar disc, the stellar bulge and the dark halo. The collisional component simulates the ISM as an ensemble of 20\,000 gas clouds that have inelastic collisions (sticky particles) that might coalesce or fragment, which evolve in the gravitational potential of the galaxy. A scheme for star formation was implemented, in which stars are formed during cloud collisions to further evolve as non-collisional particles. The star formation rate is proportional to the cloud collision rate. Ram pressure has been included as an additional acceleration on the sticky particles that are not protected by other particles. Once stripped out from the galactic plane, the warm gas clouds become diffuse if their density falls below a critical density \citep{vollmer-2021}. The model consists of two components: the HII regions ionised by young massive stars with ages less than 20 Myr and the diffuse gas ionised by the stellar UV radiation or by strong shocks induced by the ram pressure stripping process. As discussed in \citet{vollmer-2021}, collisional ionization through the thermal electrons of the ambient ICM that confines the filaments is the most probable ionization mechanism in the extraplanar ionised gas filaments detected in the VESTIGE narrow-band image.

\begin{figure*}[ht!]
\centering
\includegraphics[width=0.503\linewidth]{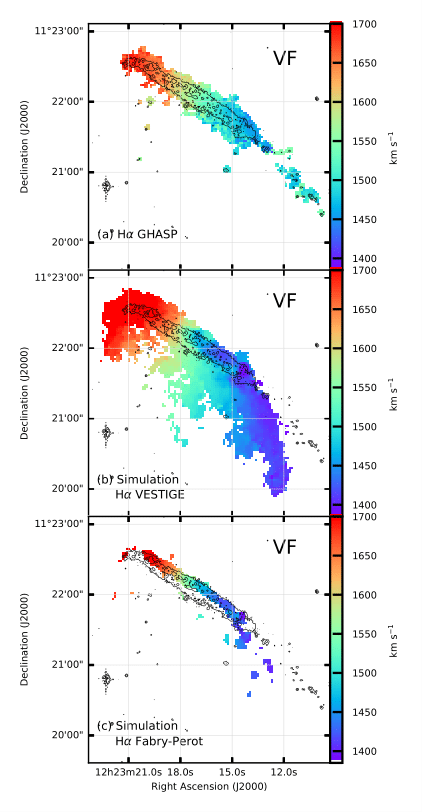}
\includegraphics[width=0.486\linewidth]{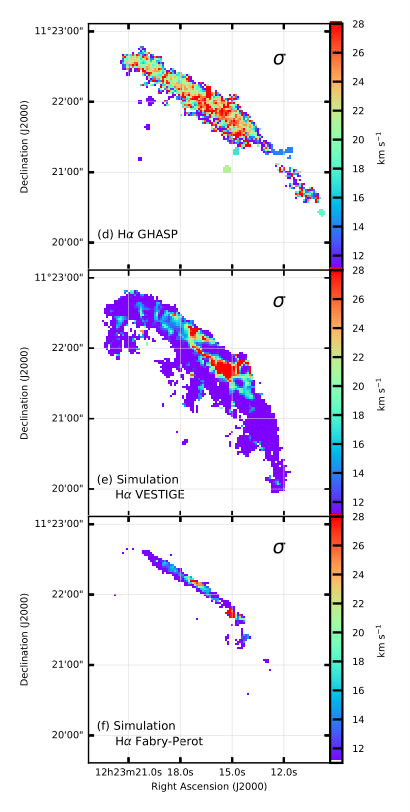}
\caption{GHASP (panel (a) and (d)) and simulated (panels (b), (c), (e) and (f)) H$\alpha$ velocity (left column) and velocity dispersion (right column) 
fields of NGC\,4330. The simulated H$\alpha$ velocity and velocity dispersion fields are given down to the faintest levels used in the middle 
panel of Fig. \ref{n4330DataModelMono} to show the kinematical properties of the low column density ionised gas reached by the simulations 
(panels (b) and (e)) and at a higher surface brightness level corresponding to the sensitivity of the FP observations 
($\Sigma(H\alpha)\simeq$ 2\,$\times$\,10$^{-17}$\,erg\,s$^{-1}$\,cm$^{-2}$\,arcsec$^{-2}$, panel (c)). The contours on the velocity fields 
indicate the H$\alpha$ surface brightness 
$\Sigma(H\alpha)\sim$\,2$\times$10$^{-18}$\,erg\,s$^{-1}$\,cm$^{-2}$\,arcsec$^{-2}$ taken from the 
VESTIGE data \citep{fossati-2018}.}
\label{n4330DataModelRV}
\end{figure*}

In Figs. \ref{n4330DataModelMono} and \ref{n4330DataModelRV}, we compare the observed H$\alpha$ data to the simulated ones.  For each figure we provide three panels, the top one corresponds to the data whereas the middle and bottom ones to the models. Figure \ref{n4330DataModelRV} (panels (b) and (e)) and \ref{n4330DataModelMono} (panel (b)) give the lowest surface brightness achieved by the model while Figs. \ref{n4330DataModelRV} (panels (c) and (f)) and \ref{n4330DataModelMono} (panel (c)) show the model at a depth comparable to the sensitivity of our observations. The plotted H$\alpha$ emission map is the one extracted from the model for the snapshots of highest goodness for the diffuse gas. This corresponds to a peak of ram pressure occured 140 Myr ago. Similarly to the deep H$\alpha$ observations presented in \cite{fossati-2018}, the model shows extra-planar linear filaments in the downwind region of the galactic disc, which correspond to the low surface brightness filaments located in the southwest tail and close to the downturn in the northeast \citep{vollmer-2021}. These filaments indicate the direction of the motion on the plane of the sky of the galaxy within the cluster.

The agreement between the gas distribution in the observations and in the simulations is overall very satisfactory. The brightest H$\alpha$ knots detected in the FP observation (panel (a)) correspond the the brightest H$\alpha$ regions in the simulation (panels (b) and (c)). The diffuse H$\alpha$ emission in the model, given (in log scale) in panel (b), is far too faint to be detected in the FP observations, but it matches fairly well the diffuse emission observed using the H$\alpha$ narrow-band filter by \cite{fossati-2018}. While the tuning is very good at the northeast edge of the disc, the simulated southwest tail has a larger curvature than that observed in the diffuse ionised gas detected by VESTIGE. This difference might be due to the fact that given the orientation of NGC\,4330 and its deduced orbit within the cluster, the northeast edge of the disc is the region of the galaxy which first suffered the stripping process.

As suggested by the simulations, all the diffuse gas component (mainly HI gas) has been already stripped, while that in the southwest direction, which was perturbed later, is still partly present, as suggested by the diffuse HI tail observed by \citet{chung-2007}. This effect is probably combined with an asymmetric distribution of the star forming regions over the disc of the galaxy, possibly formed after the external perturbation, with more compact HII regions along the southwest edge of the disc. Given their compactness and high density, these regions are less subject to the external perturbation than the diffuse gas. The difference between observations and simulations might thus result from a combined effect of sensitivity, gas density, and time elapsed since the beginning of the perturbation. Indeed, as shown in Fig. 19 of \citet{abramson-2011}, the southwest curvature of the gas is more pronounced in the cold and diffuse atomic phase (the first one stripped) than in the compact H$\alpha$, more resistant to the external pressure.

The H$\alpha$ line-of-sight velocity fields are shown in panel (a) of Fig. \ref{n4330DataModelRV}. In order to compare the model to the data, a systemic velocity of 1551\,km\,s$^{-1}$ has been added to the model. The velocity amplitudes are the same for the three panels. H$\alpha$ isocontours from \cite{fossati-2018}, corresponding to a surface brightness of $\Sigma(H\alpha)\sim$\,2\,$\times$\,10$^{-18}$\,erg\,s$^{-1}$\,cm$^{-2}$\,arcsec$^{-2}$, have been displayed in order to facilitate the comparison between the panels. We observe a very good agreement between the simulation and the observations.
Where the simulations can be compared to the data, the amplitudes of the velocity fields are consistent. We note that if we include the low surface brightness tails, unfortunately not reached by the shallow FP observations, the velocity range in the simulations is even larger.

On the other hand, the HI velocity field (Fig. \ref{hiha}, bottom panel) seems to show that the LoS velocities increase at the edge of the tail, although this rising is not very sharp, probably because of the lack of resolution in the HI data. The same excellent agreement is present also between the observed and simulated unfolded RCs shown in Fig. \ref{RC_sim_obs} on which the simulated RC measured along the major axis of the galaxy has been compared to the GHASP observed and modelled RCs.  The RC extracted from the simulations  well predict the observed
asymmetries between the receding (northeast) and approaching (southwest) sides of the disc up to $\sim$70 arcsec ($\sim$5.6 kpc).  Since the RC extracted from the simulations is taken on the major axis, we do not have any 
data outside this radius because the ionised gas emission is bent to the southwest.

The H$\alpha$ line-of-sight velocity dispersion fields are compared in the right panels of Fig. \ref{n4330DataModelRV} using the same velocity range in the three panels.  
The mean velocity dispersion in the data is $\sim$20.5\,km\,s$^{-1}$, with a standard deviation of $\sim$6.7\,km\,s$^{-1}$, while in the model those quantities are 
respectively equal to $\sim$10.6 and $\sim$6.7\,km\,s$^{-1}$.  The smaller mean velocity dispersion in the model with respect to the observations (which have an 
instrumental limiting resolution of $\sim$13\,km\,s$^{-1}$), is due to the fact that the model estimate is strongly weighted by the diffuse gas component. This component is
not reached by the observations. If we consider only the area of the model corresponding to the one detected in the data, the mean model velocity dispersion becomes $\sim$17.5 
km s$^{-1}$, with a standard deviation of $\sim$6.4\,km\,s$^{-1}$, and matches fairly well the observations. Striking is the similarity between models and observations 
around RA=$12^h23^m15.5^s$, Dec=$11^{\circ}21'42''$, where the velocity dispersion increases by one sigma with respect to other regions along the galaxy disc up to 28 - 45 
km s$^{-1}$  in the data and 30 - 40\,km\,s$^{-1}$ in the model (in red in both cases). 

\section{Discussion and conclusion} \label{section8}

Located in the nearby Virgo cluster, where 1\,arcsec corresponds to $\sim$\,80 pc, the edge-on late-type NGC\,4330 is the perfect candidate to study the effects
of ram pressure stripping in the direction perpendicular to the disc plane of a perturbed system with an angular and spectral resolution unreachable elsewhere. The presence of 
extended tails in the atomic \citep{chung-2007} and ionised \citep{fossati-2018} gas and in the dust component \citep{longobardi-2020} observed in this object
witness an ongoning stripping process able to remove the different phases of the ISM in the outer regions and producing a truncated disc.

Using two different and independent sets of FP observations with a spectral resolution up to $R\sim$\,10000, combined with our own tuned
hydrodynamic simulations \citep{vollmer-2021}, we have studied the kinematics of the 
ionised gas. Although the diffuse emission of the ionised gas in the stripped tails, which has a surface brightness of 
$\Sigma(H\alpha)$ $\simeq$ 3\,$\times$\,10$^{-18}$\,erg\,s$^{-1}$\,cm$^{-2}$\,arcsec$^{-2}$, has not been reached by the present observations,  limited to $\Sigma(H\alpha)$ $\simeq$ 10$^{-17}$\,erg\,s$^{-1}$\,cm$^{-2}$\,arcsec$^{-2}$, our data allowed us to study the kinematics of the gas 
along the disc and in a few HII regions located outside the disc plane and probably formed within the stripped gas after the interaction.

The Fourier transform analysis of the deep NGVS $i$-band image reveals that the $B_4$ parameter is negative, indicating that the galaxy disc has a boxy shape outside a 
radius of $\simeq$ 60\,arcsec ($\simeq$ 4.8\,kpc). This boxy structure is present mostly in the southwest direction, where the tail of stripped material is more pronunced.
Simulations suggests that a thicker disc could results from the perturbation of the gravitational potential well due to the displacement of the gaseous component from the
disc plane after a ram pressure stripping event \citep[][]{Farouki-1980, clarke-2017, Safarzadeh-2017ApJ...850...99S, Steyrleithner-2020MNRAS}. The boxy shape would thus result from the induced perturbation on the stellar orbits.
The datasets used in this work, which are sensitive only to the kinematics of the ionised gas, do not allow us to test this hypothesis.  
However, in the central regions, the kinematic major axis matches with the dust lane position angles, indicating that the perturbation probably had only a moderate effect on the dynamics of the inner gas.

It is thus conceivable that the dynamics of the stars are also moderately perturbed. Indeed, within the inner stellar disc ($r$ $\lesssim$ 2\,kpc), the velocity field of the galaxy dominated by the emission of HII regions is characterised by isovelocities 
parallel to the minor axis. The analysis of the rotation curves and of the PVDs consistently indicate a solid-body rotation. Outside this radius the isovelocities are bent, becoming parallel to the major axis and show a low 
velocity gradient in the southwestern edges of the disc indicating non circular motions.

The FP data allowed us to detect several ionised gas features formed after the dynamical interaction of the galaxy with the surrounding ICM (ram pressure)
and located at the edges or outside the stellar disc. These are i) a hook-like structure at the northeast of the galaxy in the region which first got in contact with the ICM,
ii) several bright and extended regions formed along the downstream direction of the wind, principally located in the southwest of the galaxy, including a few HII regions 
well outside the disc and located along the filaments of stripped gas.

The analysis of the velocity field, of the velocity dispersion, 
of the rotation curves (derived using the ETM necessary to correct for beam smearing effects due to the velocity superposition along the line-of-sight), and
of the PVDs consistently indicate that these regions do not follow the solid-body rotation but have peculiar velocities 
indicative of streaming motions out of the plane of the galaxy. We also observe a small decrease of the rotation of the stripped gas with the increasing 
distance from the galaxy disc.

This observational evidence can be explained considering that the stripping process is mainly perpendicular to the disc plane, 
as indeed suggested by our hydrodynamic simulations. 
Under this geometrical configuration, the gas is expected to keep its rotation with only a partial and gradual loss
of angular momentum. Furthermore, the similarity of the H$\alpha$, CO, and HI velocity fields and rotation curves, particularly in the inner disc, indicate that here ram pressure stripping significantly affects the gas kinematics. 
Since in this scenario the acceleration caused by ram pressure is inversely proportional to the stripped gas surface density, the observed similarity in the kinematical perturbations of the different gas components (ionised, molecular, atomic) is expected given their comparable surface density.

Overall, the kinematical properties of the galaxy and of all these extraplanar features formed after the ram pressure stripping event are well reproduced by our simulations, at least up to the limited sensitivity of our FP observations. Nevertheless, it is hard to drive general conclusions on the effects of ram pressure stripping on the kinematical properties of galaxies and of the stripped material from the
analysis of a single object.

The present results, however, combined with those obtained from the analysis of the kinematical properties of a few other well resolved galaxies 
with available IFU data \citep[e.g.][]{Chemin-2006, Merluzzi-2013MNRAS.429.1747M, Fumagalli-2014, Consolandi-2017, boselli-2021}, consistently suggest that the gas removed during a ram pressure stripping event tends to keep its 
rotation although with a possible loss of angular momentum. The perturbation can affect the kinematics of the gas even inside the disc of the perturbed system \citep[][]{2021arXiv211106635B}. Although we are still lacking of any direct evidence, indirect observations and simulations consistently indicate that a ram pressure stripping event can also perturb, although to a lower extent, the kinematics of the stellar component. Deep IFU spectroscopic observations with a sufficient spectral resolution of perturbed systems will be required to probe this scenario.

\begin{acknowledgements}
Based on observations collected at the Observatorio Astron\'omico Nacional at San Pedro M\'artir, Baja California, M\'exico (OAN -SPM). We thank the daytime and night support staff at the OAN-SPM for facilitating and helping obtain our observations. Also, based on observations taken with the GHASP and MISTRAL spectrographs at the Observatoire de Haute Provence (OHP, France), operated by the French CNRS. The authors warmly thank Olivier Boissin from LAM and the OHP team for its technical assistance before and during the observations, namely the night team: Jean Balcaen, Stéphane Favard, Jean-Pierre Troncin, Didier Gravallon and the day team led by Fran\c{c}ois Moreau as well as Dr. Auguste Le Van Suu, the Head of Observatoire de Haute Provence-Institut Pythéas. We are grateful to the whole CFHT team who assisted us in the preparation 
and in the execution of the observations and in the calibration and data reduction: Todd Burdullis, Daniel Devost, Bill Mahoney, Nadine Manset, Andreea Petric, Simon Prunet, Kanoa Withington. We acknowledge financial support from ``Programme National de Cosmologie and Galaxies" (PNCG) funded by CNRS/INSU-IN2P3-INP, CEA and CNES, France,and from ``Projet International de Coop\'eration Scientifique" (PICS) with Canada funded by the CNRS, France.
This research has made use of the NASA/IPAC Extragalactic Database (NED), which is operated by the Jet Propulsion Laboratory, California Institute of Technology, under contract with the National Aeronautics and Space Administration and of the GOLDMine database (http://goldmine.mib.infn.it/) (Gavazzi et al. 2003). M.M.S. warmly thanks the Mexican National Council on  Science and Technology (CONACyT) who found her through the program ``Becas CONACyT al Extranjero 2017", CVU 666085, and to the Secretariat of Public Education (SEP) of the Mexican Government through the scholarship ``Becas Complemento de Apoyo al Posgrado ciclo 2018-2019". M.R. thanks also the grants IN109919 of DGAPA-UNAM and CY-253085 and CF-86367 of CONACyT. MB acknowledges FONDECYT regular grant 1211000. AL is supported by Fondazione Cariplo, grant No 2018-2329. MF acknowledges funding from the European Research Council (ERC) (grant agreement No 757535).

\end{acknowledgements}

%
\bibliographystyle{aa} 
\bibliography{N4330_nov} 
%
%


\begin{appendix} 

\section{Fabry-Perot observations} 
\label{FabryPerotObservations}

\subsection{PUMA observations}
\label{PUMAobservations}

Fabry-Perot 3D spectroscopic observations of NGC\,4330 were gathered using PUMA at the 2.1 m telescope at the Observatorio Astron\'omico Nacional in San Pedro M\'artir, Baja California, M\'exico (OAN-SPM) \citep{rosado-1995}.
The PUMA focal reducer hosts a FP interferometer with a field of view of $\sim$10\,$\times$\,10 arcmin$^2$. The camera is a 512\,$\times$\,512 CCD  detector\footnote{The PUMA CCD has in fact 2048\,$\times$\,2048 px$^2$, which are electronically binned 4$\times$4 to increase the SNR per pixel by a factor 16 and to take into account the mean seeing of the site.} with a pixel scale of $\sim$1.27\,$\times$\,1.27 arcsec$^2$.

NGC\,4330 was observed during dark time, in February 2017\footnote{Additional deeper PUMA observations of NGC\,4330 could not be made due to the lockdown in spring 2020.} (see Table \ref{general}), as part of the FP survey of the \textit{Herschel} Reference Survey \citep[]{boselli-2010, gomezl-2019}. 
To increase the signal-to-noise ratio, the pixels have been electronically rebinned by 4\,$\times$\,4, leading to a final spatial sampling of $\sim$\,1.3 arcsec. 
The galaxy was observed during poor seeing conditions (FWHM\,$\sim 2.9\pm$0.3 arcsec) but excellent transparency with a   
total exposure time of 96 min, with 120 seconds per channel.
The spectral domain has been selected using a FWHM\,=\,90\,\AA\ wide filter centred at 6607\,\AA. The spectroscopic calibration of the data was made 
using the narrow Ne line at 6598.95\,\AA\ in the same interference filter. The choice of this line is optimal to minimise the phase shift effects 
due to the coating of the interferometer since it is very close to the redshifted H$\alpha$ emission line of the galaxy ($\sim$6597\,\AA).
The theoretical PUMA effective finesse is $F\sim$\,24 but the one measured using the narrow Ne spectral line was $F_e=19.9$.
The FP interference order at the mean wavelength of the observation is $p\sim328$, which gives a resolution $R=p\ F_e\sim 6533$.  
The free spectral range (FSR) of the instrument at the redshifted wavelength of the galaxy, $\sim$20.04\,\AA\ (911\,km\,s$^{-1}$), was scanned through 48 channels, corresponding to a spectral 
sampling of $\sim$\,0.42\,\AA\ (19\,km\,s$^{-1}$). The instrumental and observational parameters are listed in Table \ref{FPsetup}. 

Standard corrections were applied to the CCD  images. The astrometric calibration of the data
has been done using the \textsc{koords} task of the 
\textsc{karma}\footnote{https:www.atnf.csiro.au/computing/software/karma/}
package \citep{gooch-1996} and \textsc{wcs} (Word Coordinate System)  
textsc{IRAF}\footnote{'Image Reduction and Analysis Facility', http://iraf.noao.edu/; IRAF is distributed by the National Optical Astronomy Observatories, which are operated by the Association of Universities for Research in Astronomy, Inc., under cooperative agreement with the National Science Foundation.} task. 
The FP data were reduced and analysed using 
the \textsc{ADOCHw}
\footnote{ADHOCw, http://cesam.lam.fr/fabryperot/index/softwares, developed by J. Boulesteix.}
software, the IDL-based \textsc{ComputeEverything} 
\footnote{https://projets.lam.fr/projects/computeeverything},
\textsc{reducWizard} interface 
\footnote{https://projets.lam.fr/projects/fpreducwizard}, and our own \textsc{Python} scripts.

\begin{figure}[ht!]
\centering
\includegraphics[width=\linewidth]{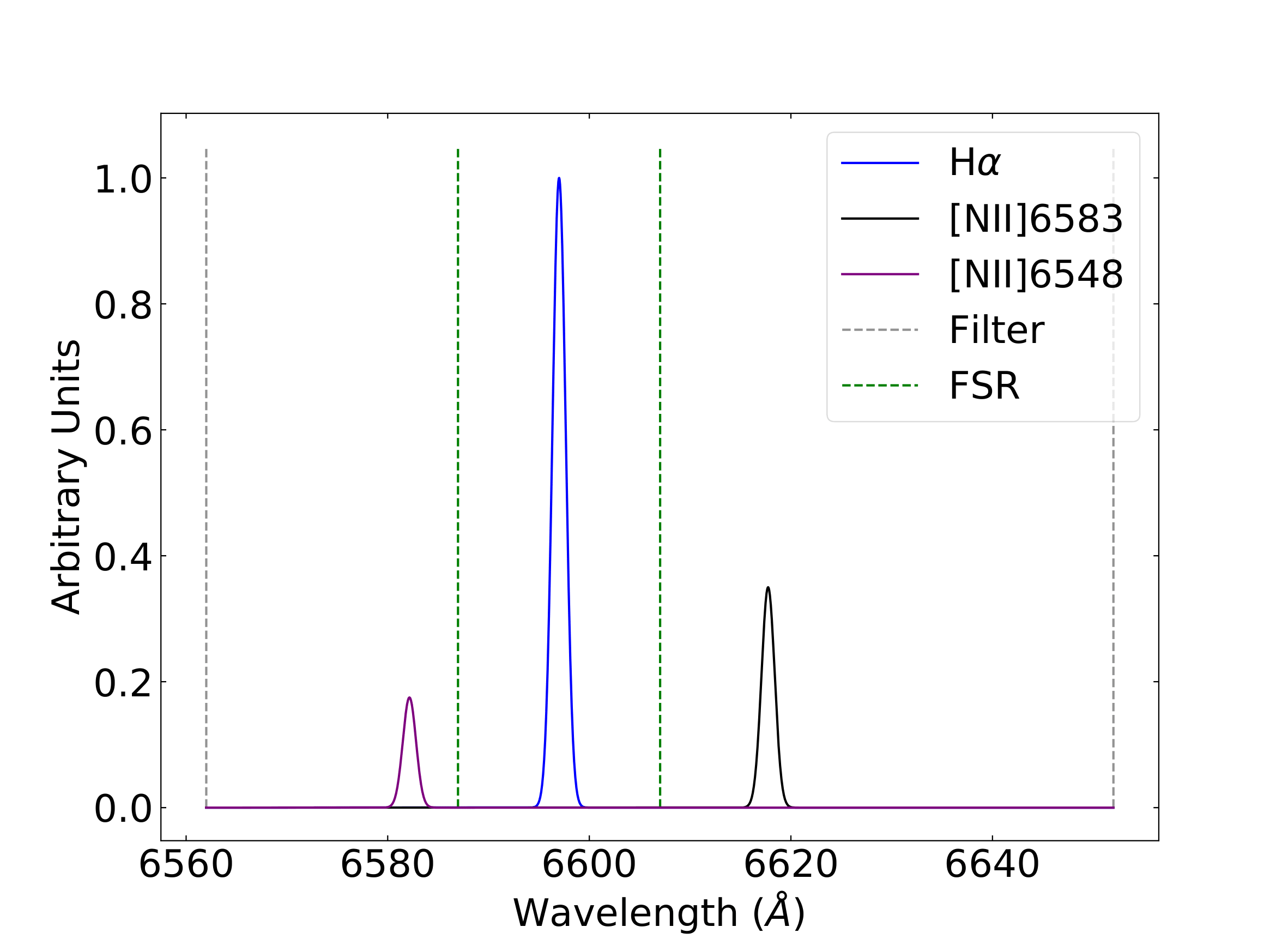}
\includegraphics[width=\linewidth]{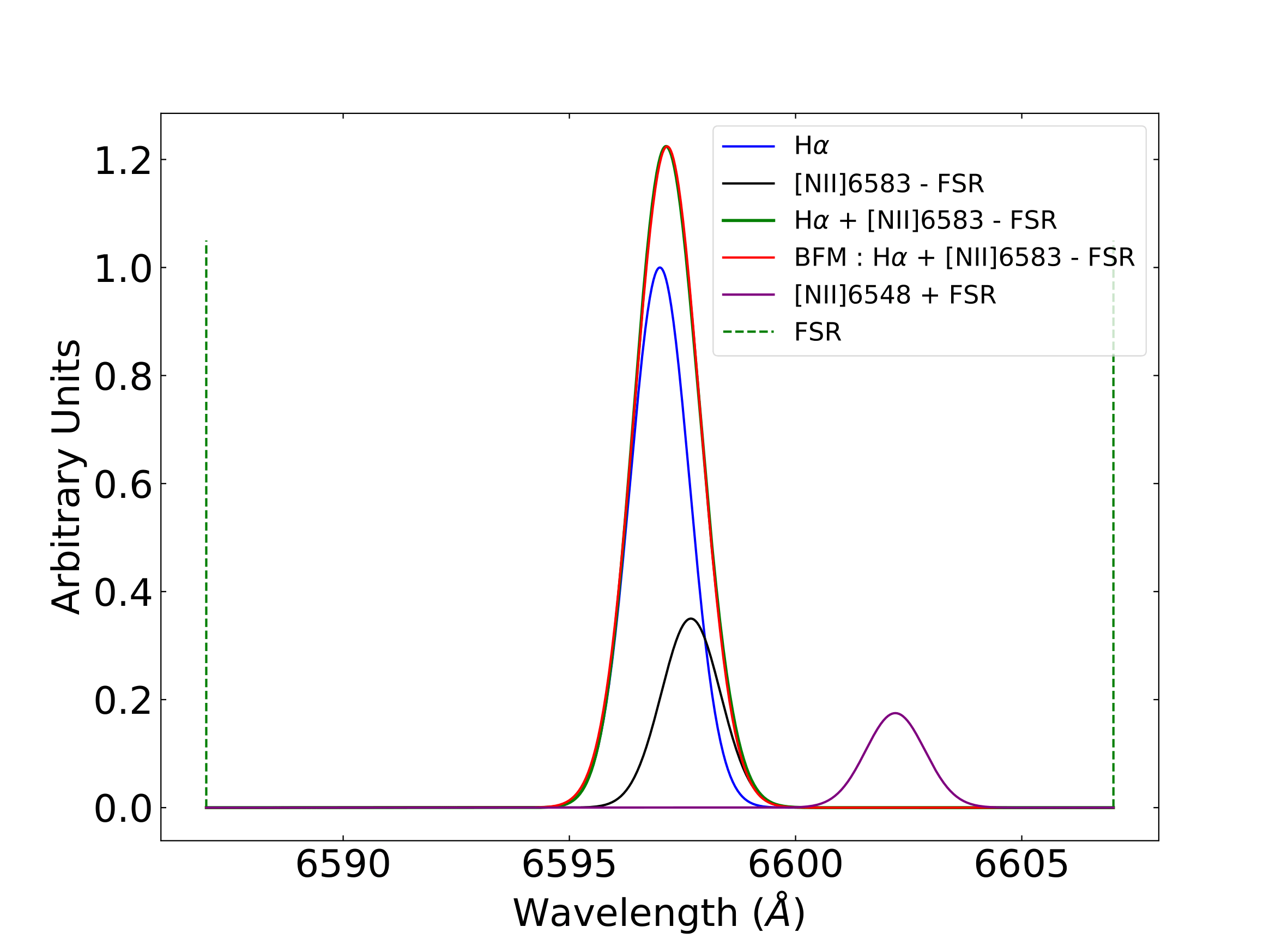}
\caption{Model of PUMA spectrum.  The H$\alpha$ plus the two [NII]6548, 6583 lines, transmitted by the filter, are represented by 
Gaussian functions of width $\sigma_{LSF}$ (see Table \ref{general}). Normalised units to the H$\alpha$ line are used for the $y$-axis.  
The relative intensity between the H$\alpha$ and the [NII] lines are given in Sect. (\ref{PUMAobservations}). The distance between the two 
green dashed vertical lines represents the Free Spectral Range (FSR\,$\sim$\,20\,\AA). Top panel: the redshifted H$\alpha$ and [NII]6548, 
6583 lines are located at their right position. The distance between the two gray dashed vertical lines represents the filter width 
(FWHM\,$\sim$\,90\,\AA). Bottom panel: the redshifted lines are located at their apparent position due to the FSR recovering: the 
[NII]6548 line is shifted by plus one FSR and the [NII]6583 is shifted by minus one FSR. The green curve represents the sum of the H$\alpha$ plus  
[NII]6583 lines whereas the red one represents the Gaussian which best fit the sum of the two lines (Best Fit Method).}
\label{FilterPuma}
\end{figure}

The filter includes the H$\alpha$ line and the [NII] lines at 6548.03 and 6583.48\,\AA\ (see Fig. \ref{FilterPuma}).  
Since the free spectral range of the interferometer is $\sim$\,4.5 times smaller than the filter width, the [NII] lines are present in the data cube but 
are respectively observed at the interference order $p+1$ and $p-1$ with respect to H$\alpha$, which is observed at order $p$.
Furthermore, the apparent separation between H$\alpha$ and [NII]6548 lines is $\sim$+5.0\,\AA\ and the one with the [NII]6583 is $\sim$+0.9\,\AA\,, as 
illustrated in the bottom panel of Fig. \ref{FilterPuma}.

Since the spectral resolution at H$\alpha$ redshifted is $\sim$1.01\,\AA, the [NII]6548 line is resolved from the H$\alpha$ line
while the [NII]6583 line partially overlaps with H$\alpha$, affecting the barycentre position and the line width measurements of the H$\alpha$ line. 
In order to correct for this effect, we make a model that takes into account (i) a constant line ratio of [NII]6583/H$\alpha$\,=\,0.35, as measured using 
the VLT/FORS long slit spectrum published in \cite{fossati-2018}, 
(ii) the known wavelength separation of $\sim$20.73\,\AA\ ($\sim$\,942\,km\,s$^{-1}$) between the redshifted H$\alpha$ and [NII]6583 lines, and (iii) 
the H$\alpha_0$ FSR at rest $\sim$19.8\AA\ ($\sim$\,905\,km\,s$^{-1}$). 
Considering that the dispersion of the instrumental line spread function is $\sigma_{LSF}\sim$\,0.43\,\AA\ ($\sim$\,19.5\,km\,s$^{-1}$), the thermal broadening $\sigma_T\sim$\,0.18\,\AA\ ($\sim$9.1\,km\,s$^{-1}$), and the intrinsic mean 
velocity dispersion $\sigma \sim$\,0.44\,\AA\ ($\sim$\,20.0\,km\,s$^{-1}$), the model indicates that the observed H$\alpha$ barycentre and the velocity dispersion
should be corrected by -0.20\,\AA\ (-8.9\,km\,s$^{-1}$) and -0.12\,\AA\ (-5.3\,km\,s$^{-1}$), respectively.

We computed a parabolic phase map from the calibration cube in order to obtain the reference wavelength for the line profile observed 
inside each pixel. This phase map provides the shift that has to be applied in the spectral dimension to every pixel of the interferogram 
cube to bring all channels to the same wavelength. 
The wavelength-sorted data cube is then created by applying the phase map correction to the interferogram data cube. 
A detailed explanation about the data reduction process for data obtained at OAN-SPM with PUMA can be found in \cite{fuentesc-2004}. 

In order to work with profiles and maps with a similar signal-to-noise ratio (SNR) and at the same time optimise the angular resolution, we applied a Voronoi tessellation with a constant 
$\mathrm{SNR}=9$ to the data cube \cite[][]{capellari-2003, daigle-2006-a, daigle-2006-b}. The brightest HII regions are not affected by this smoothing procedure because their SNR per pixel is larger than 9 and are thus displayed at full resolution.

The resulting data cubes are used to extract the monochromatic, continuum, line-of-sight (LoS) radial velocity and velocity dispersion maps of the galaxy. 
The continuum map is computed considering the average of the 3 lowest intensities of the 48 channels of the cube. For the monochromatic 
image, the intensity of the H$\alpha$ line is obtained by integrating the flux of the line profile for each pixel. 
Following \cite{daigle-2006-a}, we calculate the radial velocity in each pixel by measuring the barycentre of the profile 
of the H$\alpha$ line. This barycentre is computed as the intensity weighted centroid of the spectral bins falling into 
the emission line boundaries, with the continuum subtracted. Radial velocities and velocity dispersions are extracted using a single 
emission-line detection algorithm. In the case where more than one velocity component is present in the spectrum, only the strongest 
emission line is taken into account. When two emission lines are spectrally close and have comparable amplitudes, they might be
taken as a single one with a larger velocity dispersion. 

Assuming that all the H$\alpha$ emission-profiles are described by Gaussian functions, from the emission profiles width  map the velocity dispersion $\sigma$ was estimated as
\begin{equation}
\sigma=(\sigma^{2}_{\rm{LoS}}-\sigma^{2}_{\rm{LSF}}-\sigma^{2}_T)^{1/2},
\label{sigmaLoS}
\end{equation} 
where $\sigma_{\rm{LoS}}$ is the observed LoS velocity dispersion directly measured from the data cube corrected (i) from the instrumental line-spread function \citep[LSF, e.g.][]{valdezg-2001, Rosado-2013, cardenas-2018}, which is modelled using the instrumental finesse F and the free spectral range FSR by $\sigma_{\rm{LSF}}=\mathrm{FSR}/(\mathrm{F}\times 2\sqrt{2\mathrm{ln} 2})\sim$19.5 $\mathrm{\,km\,s^{-1}}$ (see Table \ref{FPsetup}) and (ii) from the thermal broadening $\sigma_{T}=(kT_{e}/m_{H})^{1/2}=9.1\,\,\,\mathrm{km\,s^{-1}}$, assuming an electronic temperature of HII regions of $T_{e}=10^{4}\,\,\,\mathrm{K}$ \citep{osterbrock-2006}.

\label{PUMA observations}

\subsection{GHASP observations}
\label{GHASPobservations}

An independent set of FP spectroscopic observations was taken using the GHASP instrument on the 1.93 m telescope at the Observatoire de Haute 
Provence (OHP) in spring 2021. These observations were gathered using the same configuration extensively described in \cite{gomezl-2019}. 
The GHASP focal reducer is equipped with a FP with a field of view of 5.8\,$\times$\,5.8 arcmin$^2$ and a 512\,$\times$\,512 Imaging Photon 
Counting System\footnote{The GHASP IPCS has in fact 1024\,$\times$\,1024 px$^2$, which are electronically binned 2$\times$2 to increase by a factor 4 the frame readout frequency and to take into account the mean seeing of the site.  Let us remember that for an IPCS the SNR does not increase with the pixel binning.} with a pixel scale of 0.68\,$\times$\,0.68 arcsec$^2$ \citep{gach-2002}. The observations were taken during poor seeing conditions 
($\mathrm{2.5<FWHM<4.0}$\,arcsec), with medium transparent sky, and with a total integration time of 645 minutes. The FSR of the FP at the redshifted wavelength of the galaxy
(379\,km\,s$^{-1}$) was scanned through 32 channels, with a typical spectral resolution of $R\sim$\,9704 at H$\alpha$ redshifted. 
NGC\,4330 was observed using two different narrow-band interference filters. Indeed,  the mean redshifted H$\alpha$ wavelength of the galaxy is 
$\sim$\,6597\,\AA, which unfortunately falls just in the middle of two available interference filters centred on 6591\,\AA\ and 6601\,\AA. Both  
filters have the same FWHM\,$\sim$15\,\AA\, useful to overcome [NII] line contamination.   
To cover the whole velocity field of the galaxy, the first filter was used for the blueshifted part of the galaxy,
the second one for the redshifted side. This does not exactly double the observing time because approximatively the two third of the velocity 
amplitude was covered by the two filters that overlap by $\sim$5\,\AA. Similarly to the PUMA observation, the wavelength calibrations were done 
using the NeI emission line at $\lambda$\,=\,6598.95\,\AA. The data reduction procedure adopted to reduce the GHASP data have been extensively 
described in \cite{epinat-2008-a} and \cite{gomezl-2019}. The LoS velocity dispersion has been corrected from the line spread function (LSF) and thermal 
broadening using the method and the relations described in the previous section (Sect. \ref{PUMA observations}). The main observational parameters of the 
FP observations are summarised in Table \ref{FPsetup}.

We decided not to combine the FP observations obtained with GHASP and PUMA because of their different spatial and spectral resolutions.
On the contrary, we decided to keep them separated to i) check the consistency of the results in particular in the
low surface brightness regions where the signal-to-noise is limited, ii) take benefit of the slightly better angular resolution of the PUMA data
for the study of the extraplanar gas, and iii) of the higher spectral resolution of the GHASP data for the study of the velocity dispersion of 
the gas, and finally iv) use the deeper GHASP data for the comparison with the output of the simulations.
We notice that for these purposes the observations were analysed differently, i.e. using a Voronoi adaptive spatial smoothing for the PUMA data 
in order not to thicken the almost edge-on disc of the galaxy, while using a Gaussian smoothing for the GHASP observations to detect the weakest structures.

The different data cubes and maps presented in the next sections were thresholded using the VESTIGE image; all pixels below a surface brightness limit of 1 
and 3.5\,$\times$\,10$^{-17}$ $\mathrm{erg\ s^{-1}\ cm^{-2}\, arcsec^{-2}}$, for
GHASP and PUMA respectively, have been masked. Nevertheless, because of the night sky lines contamination, for the GHASP data, even above this threshold 
limit,  it was not possible to calculate a consistent line flux nor velocity, and velocity dispersion for each pixel in the regions labelled B, C, D, H, J 
(western tail only) and M. For each region, the profiles were summed to give average values, which in turn lead to average flux, LoS and 
velocity dispersions.

\begin{table*}
\caption{Set-up of the Fabry-Perot observations}
\begin{center}
\begin{tabular}{lcccc}
\hline\hline
Parameter 							& OAN-SPM	& 				& \multicolumn{2}{c}{OHP}	 \\ 
\hline
Telescope 							& 2.1\,m  		& 			& \multicolumn{2}{c}{1.93\,m}	\\ 
Instrument 							& PUMA$^{(1)}$ 	&    	    & \multicolumn{2}{c}{GHASP}		\\ 
Seeing 						& 2.9\,arcsec   &			& \multicolumn{2}{c}{2.5 - 4.5\,arcsec }	\\ 
Galaxy H$\alpha$ wavelength range & 			& 6593 - 6600 \AA 		& \multicolumn{2}{c}{}	 	\\ 
\,\,\,\,\,\,\,\,\,\,\,\,\,\,\,\,\,\,\,\,\,\,\,\,\,\,\,\,velocity range & 			& 1380 - 1672 $\mathrm{km\,s^{-1}}$ 		& \multicolumn{2}{c}{}	 	\\ 
Observation date 	& 2017-02-23	&  & \multicolumn{1}{c}{ 2021-01-14 to 18}	&  \multicolumn{1}{c}{2021-01-16 to 18}\\ 

& & &\multicolumn{1}{c}{ 2021-02-11} & \\
Number of cycles 						& 1 			& 					& \multicolumn{1}{c}{80}			&  \multicolumn{1}{c}{37}\\ 
Total exposure time 					& 96\,min 			& 					& \multicolumn{1}{c}{	427\,min }			&  \multicolumn{1}{c}{200\,min }\\ 
\hline
Scanning FP interferometer 					& 			& ICOS - ET-50$^{(2)}$			& \multicolumn{2}{c}{}		 	\\ 
Calibration (Neon line)					& 			& 6598.95\,\AA				& \multicolumn{2}{c}{}	\\ 
Finesse (F)						& 19.9 			& 					& \multicolumn{2}{c}{	12.3$\pm$0.3$^{(3)}$} \\
p$^{(4)}$ at H$\alpha_0^{(5)}$=6562.8 (H$\alpha^{(5)}$= 6596.8\AA) 	& 330 (328) 		& 					& \multicolumn{2}{c}{793 (789)}	\\
Resolution power (pF) at H$\alpha^{(5)}$				&  6533			& 					& \multicolumn{2}{c}{	9704$\pm$25}	\\ 
Free spectral range at H$\alpha^{(5)}$ 	&  20.05\,\AA\,/\,911\,$\mathrm{km\,s^{-1}}$		& 					& \multicolumn{2}{c}{	8.35\,\AA\, / 379\,$\mathrm{km\,s^{-1}}$}	\\ 
Spectral resolution at H$\alpha^{(5)}$  &   1.01\,\AA\,/\,45.8\,$\mathrm{km\,s^{-1}}$ 		& 					& \multicolumn{2}{c}{	0.68\,\AA\,/\,30.9\,$\mathrm{km\,s^{-1}}$ }	 \\ 
Spectral sampling at H$\alpha^{(5)}$ 	&  0.42\,\AA\,/\,19.0\,$\mathrm{km\,s^{-1}}$		& 					& \multicolumn{2}{c}{	0.26\,\AA\,/\,11.9\,$\mathrm{km\,s^{-1}}$}	\\
Line Spread Function ($\sigma_{LSF}$) &  0.43\,\AA\,/\,19.5\,$\mathrm{km\,s^{-1}}$ 	& 					& \multicolumn{2}{c}{	0.29\,\AA\,/\,13.1\,$\mathrm{km\,s^{-1}}$}	 \\  
Scanning steps 							& 48  			&					& \multicolumn{2}{c}{	32}	\\ 
Exposure time per channel			& 120\,s  			&					& \multicolumn{2}{c}{10\,s}	\\ 
\hline
Detector Name 							& Spectral 2 $^{(6)}$	&					& \multicolumn{2}{c}{IPCS (GaAs)}	 	\\
Detector size 					& 512 $\times$ 512\,pixels$^2$ 	&					& \multicolumn{2}{c}{512 $\times$ 512\,pixels$^2$}	  	\\ 
Field-of-view 				& 10 $\times$ 10\,arcmin$^2$ 	&					& \multicolumn{2}{c}{5.8 $\times$ 5.8\,arcmin$^2$ }	  	\\ 
Image scale 				& 1.27$^{(7)}$\,arcsec\,pix$^{-1}$ 		&					& \multicolumn{2}{c}{0.68$^{(8)}$\,arcsec\,pix$^{-1}$ }	 	\\ 
\hline
\end{tabular}
\label{FPsetup}
\end{center}
\tablefoot{
$^{(1)}$ https://www.astrossp.unam.mx/instrumentos/interferometria/puma/docs/manualespuma/puma.html. 
$^{(2)}$ \textit{Queensgate Instruments}, which is now called \textit{IC Optical Systems}. 
$^{(3)}$ Mean and standard deviation of the finesse ($F$) measured from eight different calibration cubes. 
$^{(4)}$ p is the interference order. 
$^{(5)}$ H$\alpha_0$ is the wavelength of the H$\alpha$-line at rest, and H$\alpha$ the wavelength at the galaxy heliocentric line-of-sight velocity (see Table \ref{general}). 
$^{(6)}$ https://www.astrossp.unam.mx/en/users/ccd-s. 
$^{(7)}$ After 4 $\times$4 electronic binning on the CCD. %
$^{(8)}$ After 2 $\times$2 electronic binning on the IPCS. %
}
\label{observations}
\end{table*}

\section{Derivation of the rotation curve}
\label{DerivationOfTheRotationCurve}

\subsection{Position velocity diagrams}
\label{Sec:pvds}

In order to study the velocity distribution as a function of the H$\alpha$ surface brightness, we built position-velocity  diagrams (PVDs) from the FP data cube after having subtracted the stellar continuum \citep[e.g.][]{epinat-2008-a, Rosado-2013}. Using a pseudo-slit of 2 pixels width (i.e. $\sim$2.6/1.4 arcsec for PUMA/GHASP, close to or below the seeing) centred at the photometric centre and aligned to the major axis defined in Section \ref{section3}, we extracted the emission intensity projected along the major axis of the galaxy after rotating the data cube in order to align the position angle of the galaxy  with the North-South direction (see Figure \ref{Figure-PVDs}). We modified the $PA$ and the centre of the pseudo-slit in order to obtain the most symmetric PVD in intensity, the final values correspond to the kinematic major axis ($PA_{kin}$) and kinematic centre (see Table  \ref{pumavsghasp}).

\subsection{The intensity-peak method}

The intensity-peak method is based either on a barycentric measurement of the line profiles for each pixel or by fitting it with a theoretical function such as a 
Gaussian \citep[e.g.][]{sofue-2001, Rosado-2013}.
With this method, the tangential velocity in the galactic plane, $V_{rot}$, is computed under the assumption that the galaxy has an  infinitesimally flat disc, inclined by an angle $i$ with respect to the sky plane and dominated by rotational motions around an axis perpendicular to the galactic plane.  Assuming an axial symmetry about the galactic centre and considering that the velocity field is dominated by circular motions, we can measure the velocity of the pixels that are at the same distance from the kinematic centre along the major axis by dividing the galaxy on the sky-plane in rings. The rotation velocity
$V_{rot}(r)$ for a ring of radius $r$ is given as the azimutal average over the ring of the deprojected velocities:
\begin{equation}
V_{rot}(r, \theta) = \frac{V_{\rm{LoS}}(r, \theta)-V_{sys}}{\cos{\theta} \sin{i}}
\end{equation}
where $r$ and $ \theta$ are the radial and angular coordinates in the plane of the galaxy \citep{mihalas} and $V_{\rm{LoS}}(r,\, \theta)$ is the line-of-sight velocity extracted from each pixel of the FP velocity field \citep[e.g.][]{fuentesc-2004, epinat-2008-b, cardenas-2018, sardaneta-2020}.

This method has been used for the PUMA and GHASP datasets; the results are shown by the black filled and open squares in Fig. \ref{RCET}. In both cases, the RCs were computed 
considering all the pixels located within an angular sector of $\pm$5$^{\circ}$ from the kinematic major axis, $V_{rot}$ have been computed within circular rings of 6.5\,arcsecond a galaxy kinematical inclination $i=88\degr$. 
Both sides of each RC show a slowly rising solid-body shape despite the fact that both sides do not match outside the inner radius $r_{0}\sim$30 
arcsec ($\sim$2.4\,kpc), but the increasing trend of the curve continue up to $\sim$95\,arcsec ($\sim$8\,kpc). 
We have indicated $r_{0}$ by dashed lines in all the maps and PV diagrams all over the paper. Beyond this radius, differences are observed between the 
receding and approaching sides. Discrepant points are associated to several regions previously identified in Fig. \ref{Hacomposite}: region A on the 
hook-like structure and region J located at about the same galacto-centric distance show a velocity discrepancy 
of $\sim$110\,km\,s$^{-1}$ in the GHASP RC and $\sim$50\,km\,s$^{-1}$ in the PUMA RC; regions K and L in the southwest tail do not have a counterpart 
on the other side of the galaxy.  The deepest GHASP data show velocities fairly constant ($V_{rot}$ $\sim$ 122 $\pm$ 10\,km\,s$^{-1}$) in the southwest tail, 
consistent with a streaming motion.

\subsection{The envelope-tracing method}
\label{The envelope-tracing method}

The envelope-tracing method (ETM) uses the position-velocity diagram (PVD) traced along the major axis of the galaxy in simulating a pseudo-slit having a width equal to two pixels (see Sect. \ref{Sec:pvds}) and adopts all the kinematic parameters (centre, 
position angle, and systemic velocity, see Table \ref{pumavsghasp}) derived from it, with the exception of the inclination which cannot be determined using this method. Following \cite{sofue-1997AJ....114.2428S, sofue-1999PASJ...51..737S,garcia-ruiz-2002}, 
the rotation velocity is defined as:
\begin{equation}
V_{rot}\, \sin i = (V_{env}-V_{sys}) - (\sigma_{T}^{2}+\sigma_{LSF}^{2})^{1/2},
\label{sofue-ec3}
\end{equation}

where $V_{sys}$ is the systemic velocity, $i$ is the galaxy inclination with respect to the sky plane, $\sigma_{LSF}$ is the instrumental 
velocity dispersion, $\sigma_{T}$ 
is the thermal broadening of the gas, and $V_{env}$ is the terminal velocity characterised by the intensity of the envelope of an observed PVD along the line profile\footnote{\cite{sofue-2001} provide a slightly different expression that corrects for instrumental and thermal broadening after deprojecting from inclination. Despite their work is often cited as reference since then, we believe this was a typo since the term that needs to be corrected for broadening is the observable, $V_{env}-V_{sys}$, as done in the previous work of \cite{sofue-1997AJ....114.2428S}.}. 
The intensity of the envelope is defined as:
\begin{equation}
I_{env}=[(\eta I_{max})^{2}+(I_{min})^{2}]^{1/2},
\label{sofue-ec-intensity}
\end{equation}
where $I_{max}$ is the maximum intensity in the line profile, $I_{min}$ is a minimum value of the contour, typically taken to be 3\,$\times$\,$rms$ noise in the PVD, and 
$\eta$ is a constant that determines the fraction of maximum intensity to set the envelope intensity, that is usually taken in the range $0.2-0.5$ (see references above). An arc-tangent model reproduces fairly well the shape of the envelope of the PVD with the smallest number of arguments and provides an adequate match to most rotation curves \citep{Courteau-1997AJ....114.2402C}. 
This function is given by:
\begin{equation}
V_{env}(r)= V_{sys} + V_{0}\, \tan^{-1} \left(\frac{r-r_{0}}{r_t}\right) + C
\label{eq-vr-teorica}
\end{equation}
where $V_{sys}$, $V_{0}$, $r_{0}$, $r_{t}$ and $C$ are five free parameters:
$V_{sys}$ is the systemic velocity, $V_{0}$ is the velocity amplitude with respect to the velocity $C$ at $r=r_0$, $r_0$ and $r_{t}$ are two characteristic radii that shape the curve marking the radius between the rising and flat part of the RC and the sharpness of this transition respectively.
$r_{0}$ is indeed the inflexion point of the function visible on the pink curve on Fig. \ref{RCET}, and represented by two dashed lines on the H$\alpha$ velocity fields (Fig. \ref{velocityfields}). This arc-tangent model has also been used by other authors \cite[][]{Weiner-2006ApJ...653.1027W, Drew-2018ApJ...869...58D, Zhao-2021ApJ...913..111Z}.
Because the PVD is traced along the major axis, relation (\ref{eq-vr-teorica}) can be transformed into a rotation curve using relation (\ref{sofue-ec3}).
Therefore, to apply the ETM,  we determined the terminal velocity $V_{env}$ by fitting the arc-tangent model to both the maximum $I_{max}$ and minimum $I_{min}$ intensity 
line profiles. We corrected the velocities for $\sigma_{LSF}$ for both instruments (see Table \ref{pumavsghasp}) and for $\sigma_{T}=9.1\,\mathrm{km\,s^{-1}}$ to account for 
the dispersion due to the gas temperature along the LoS (see Sect. \ref{section2}).
The results of the ETM are shown in Fig. \ref{RCET}. The rotation velocities and its best fit are given on the upper panels (a) and (c) while the PVDs from which the rotation velocities 
have been computed are shown on the lowest panel (b) and (d). 
Using Eq. (\ref{sofue-ec-intensity}), we computed the intensity of the envelope  ($I_{env}$) for $\eta$=[0.2-0.5] and then we applied relation (\ref{sofue-ec3}).  
The shaded area corresponds to $\eta$=[0.2-0.5] and the pink curve to $\eta$=0.3.  For $\eta$=0.3, the 
maximum coefficient of determination is $R^{2}=0.8$ and was obtained with the parameters given in Table  \ref{pumavsghasp}. 
\begin{table*}[ht!]
\caption{Photometric and kinematic parameters}
\begin{center}
\begin{tabular}{cccc}
\hline\hline
Parameter & Photometric & \multicolumn{2}{c}{Kinematic} \\ 
 &  & PUMA & GHASP \\ 
 \hline
 \rule[-0.2cm]{0cm}{0cm} $C_\alpha$ (J2000)$^{1}$ & $12^{h}23^{m}17.1^{s}$ & $12^{h}23^{m}17.0^{s}$ & $12^{h}23^{m}17.0^{s}$ \\ 
 \rule[-0.2cm]{0cm}{0cm} $C_\delta$   (J2000)$^{1}$& $+11^{\circ}\,22'\,5.7$\arcsec  &  $+11^{\circ}22'06.7$\arcsec  & $+11^{\circ}22'06.7$\arcsec \\ 
 \rule[-0.2cm]{0cm}{0cm} PA (deg)$^{2}$ &  $59.0 \pm 0.1$  & $56\pm1$ & $ 59\pm1$  \\ 
 \rule[-0.2cm]{0cm}{0cm} i (deg)$^{3}$ & $89.5 \pm 0.1$  & $ 88\pm2$  & $ 88\pm2$  \\ 
 \rule[-0.2cm]{0cm}{0cm} $V_{sys}$ ($\mathrm{km\, s^{-1}}$)$^{4}$ & ...    & $1569\pm2$ & $1563\pm2$ \\ 
 \rule[-0.2cm]{0cm}{0cm} $V_{rot}^{ETM}$  ($\mathrm{km\, s^{-1}}$)$^{5}$ & ...    & $ 164\pm11$  & $ 158\pm8$  \\ 
 \rule[-0.2cm]{0cm}{0cm} $V_{rot}^{ITM}$ ($\mathrm{km\, s^{-1}}$)$^{6}$ & ...    & $162\pm8$ & $150\pm21$ \\ 
 \rule[-0.2cm]{0cm}{0cm} $\langle \sigma_{obs} \rangle$ ($\mathrm{km\, s^{-1}}$)$^{7}$ &  ...   & $ 35\pm10$  & $ 21\pm7$  \\ 
 \rule[-0.2cm]{0cm}{0cm} $\sigma_{LSF}$ ($\mathrm{km\, s^{-1}}$)$^{8}$ & ...   & 19.5 & 13.1 \\
 \rule[-0.2cm]{0cm}{0cm} $V_{0}^{ETM}$ ($\mathrm{km\, s^{-1}}$)$^{9}$ & ...   & $46\pm8$ & $34\pm5$ \\
 \rule[-0.2cm]{0cm}{0cm} $r_{0}^{ETM}$ (arcsec)$^{10}$ & ...   & $27\pm2$ & $28\pm2$ \\
 \rule[-0.2cm]{0cm}{0cm} $r_{t}^{ETM}$  (arcsec)$^{11}$ & ...   & $17\pm4$ & $15\pm4$ \\
 \rule[-0.2cm]{0cm}{0cm} $C^{ETM}$ ($\mathrm{km\, s^{-1}}$)$^{12}$ & ...   & $98\pm3$ & $100\pm2$\\
 \rule[-0.2cm]{0cm}{0cm} $V_{0}^{ITM}$ ($\mathrm{km\, s^{-1}}$)$^{13}$ & ...   & $107\pm5$ & $110\pm15$ \\
$r_{t}^{ITM}$ (arcsec)$^{14}$ & ...   & $8\pm3$ & $20\pm1$\\
\hline
\end{tabular}
\end{center}
\label{pumavsghasp}
\tablefoot{
(1): Right ascension and declination of the galaxy centre.  The photometric and the kinematic centre differ by less than half the seeing disc.
(2): Position angle of the major axis.
(3): Column 2: photometric inclination computed with the \textsc{ellipse} task from \textsc{IRAF}. Columns 3 and 4: kinematic inclination computed with the iteration method (ITM).
(4): Systemic velocity.
(5): Rotation velocity measured with the RC computed with the envelope-tracing method (ETM) at $R_{23.5}$.
(6): Rotation velocity measured with the RC computed with the iteration method (ITM) at $R_{23.5}$.
(7): Mean velocity dispersion.
(8): Instrumental velocity dispersion (also called the line-spread function, LSF).
(9) to (14):  Best fit parameters of equation \ref{eq-vr-teorica} to compute the RC with the ETM (see text).}
\end{table*}
\subsection{The iteration method}
\label{ITM}
The intensity-peak and the envelope-tracing methods are limited by the observational parameters such as the seeing, the spectral resolution, the slit width (or pseudo-slit width in case of 2D velocity fields). They also depend on the physical properties of the galaxy such as its inclination, the velocity dispersion of the gas and its distribution within the disc. To overcome these sources of uncertainties, \cite{takamiya-2000} proposed a different model, the iteration method (ITM), to generate a model data cube using specific radial and vertical density profiles of the ionised gas (e.g. exponentially declining), rotation curves, velocity dispersions, and viewing angles.
First, to provide the initial conditions from which a first model data cube is computed, we used a rotation curve derived from a position-velocity diagram. Second, the ETM is used to derive a new rotation curve from the first model data cube, which is compared to the data.  Third, the differences between these two RCs are added to the input rotation curve to generate a new model. This procedure is repeated until the difference between the rotation curves derived from the data and the final model satisfies a convergence criterion or until a maximum number of iterations has been performed \citep[e.g.][]{takamiya-2000, sofue-2001, heald-2006, heald-2007}. 
As suggested by the intensity-weighted and the envelope-tracing rotation curves of NGC\,4330, the inner regions of the disc within $R_{23}$ show a solid-body shape (see Fig. \ref{RCET} and \ref{velocityfields}).

Therefore, following \cite{swaters-1997, heald-2006, Rosado-2013} and \cite{moiseev-2014}, we applied to NGC\,4330 the classically called cylindrical model. This model assumes, as a first grade approximation, that the galaxy has the shape of a cylinder  that rotates as a rigid body around the $z$-axis perpendicular to the $x,y$-plane, which defines the disc.
In order to compute the model data cube, we proposed an initial rotation curve that describes the rotation of the cylindrical shaped rigid body based on the arc-tangent 
model of a rotation curve (Eq. \ref{eq-vr-teorica}). Using the non-linear least-squares minimization of the library \textsc{LmFit} from \textsc{Python}, we computed a model 
data cube for each FP data set whose emission profiles were represented by Gaussian functions. The velocity dispersion of the emission peaks was fixed for all profiles 
to the medium value of the velocity dispersion of the galaxy $\langle\sigma_{\rm{LoS}}\rangle$.
For the amplitude of the Gaussian functions, we assumed that the gas has an exponentially declining distribution \citep[][]{mo, takamiya-2002} within the disc of the galaxy along the $z$-direction, $B(z)=B_{0}\exp(-|z|/z_{d})$, and we measured the vertical disc length $z_d$\,=\,6\,arcsec using the H$\alpha$ surface brightness distribution. We used the value of the kinematic centre of each data set as the systemic velocity of the galaxy ($V_{sys}$) and the kinematic $PA$ obtained from the PVDs. This implies that the free parameters of the model are: the transition radius between the rising and flat part of the RC ($r_{t}$), the  asymptotic velocity ($V_{0}$), and the inclination ($i$) of the galaxy.

\section{Additional figures}
\label{appendix:addional-figures}

\begin{figure}[ht!]
\centering
\includegraphics[width=\linewidth]{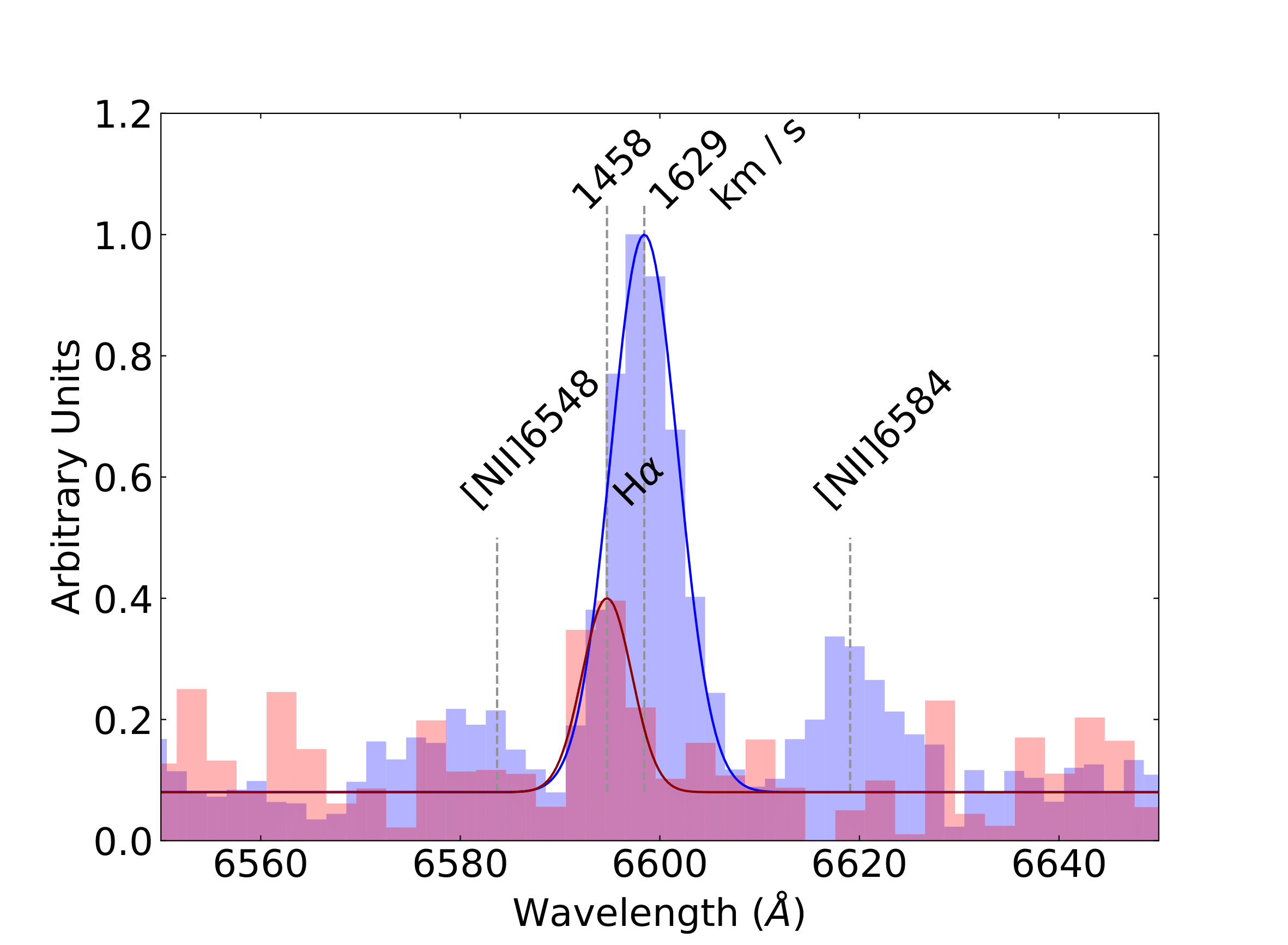}
\caption{Long-slit spectra obtained with MISTRAL at the 1.93 m OHP telescope. The H$\alpha$ lines of region D (red spectrum) and the position of the orthogonal projection of region D on the major axis of the galaxy (blue spectrum) are fitted by Gaussian functions, with intensities in normalised units.  The two vertical dashed lines on H$\alpha$ indicate the measured velocity of region D (1458$\pm$46\,km\,s$^{-1}$) 
and on the disc (1629$\pm$46\,km\,s$^{-1}$).}
\label{mistral}
\end{figure}
\begin{figure*}[ht!]
\centering
\includegraphics[width=\linewidth]{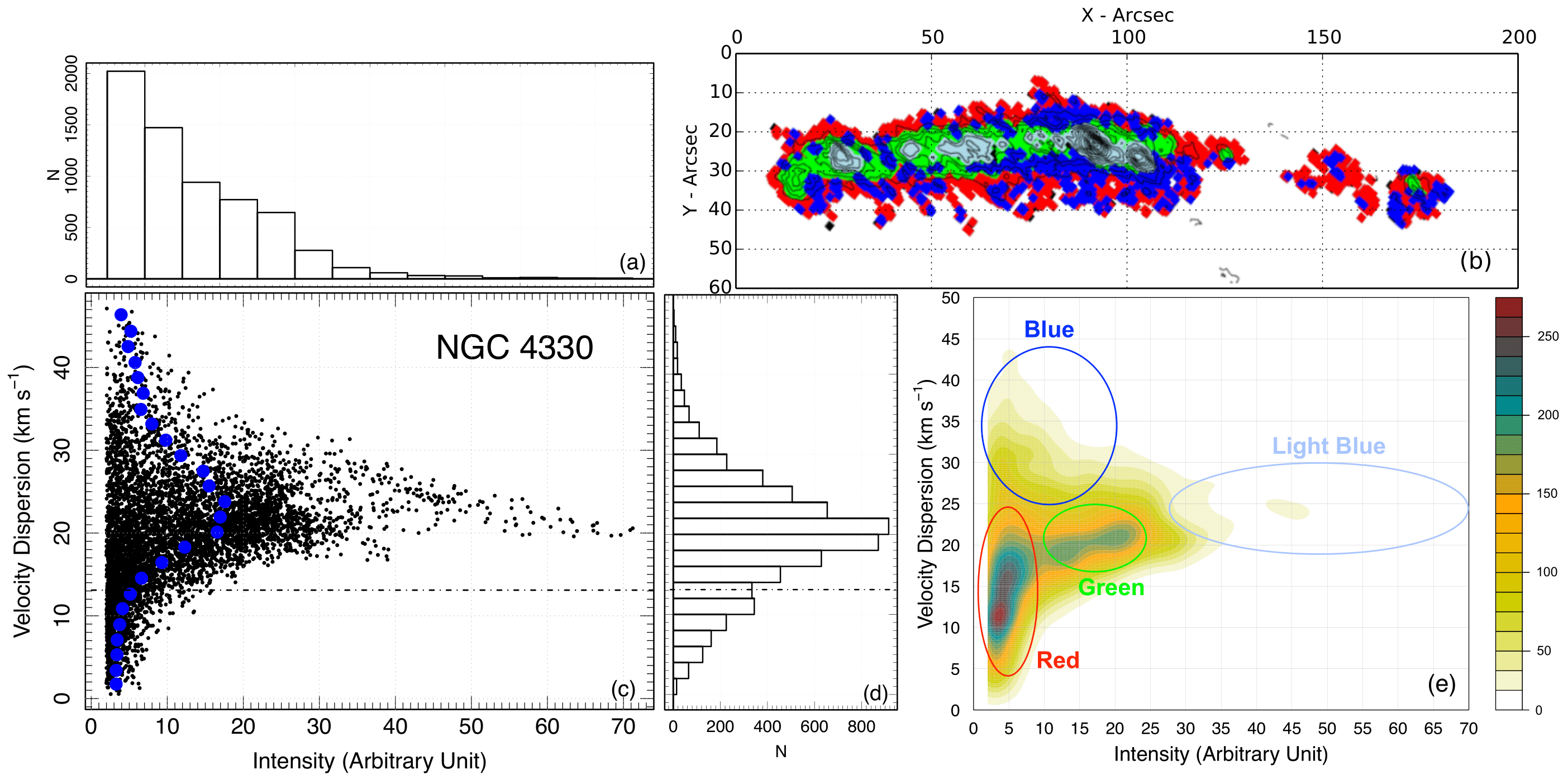}
\caption{GHASP velocity dispersion properties of NGC\,4330 :
(a) intensity histogram;
(b) loci, in the galaxy, of the different areas shown in panel (e) with different colors (red, green, blue and light blue) on top of which H$\alpha$ isocontours have been superimposed; 
(c) velocity dispersion vs. emission intensity diagnostic diagram, each black dot represents a pixel and the blue dots represent the mean of the intensity in each velocity dispersion bin;
(d) velocity dispersion histogram;
(e) pixel density map of the velocity dispersion vs. intensity diagram, the color bar represents pixel density.
The dashed line  in plots (c) and (d) represents the Line Spread Function $\sigma_{LSF}$.} 
\label{henri}
\end{figure*}

\begin{figure*}[ht!] 
\includegraphics[width=1\linewidth]{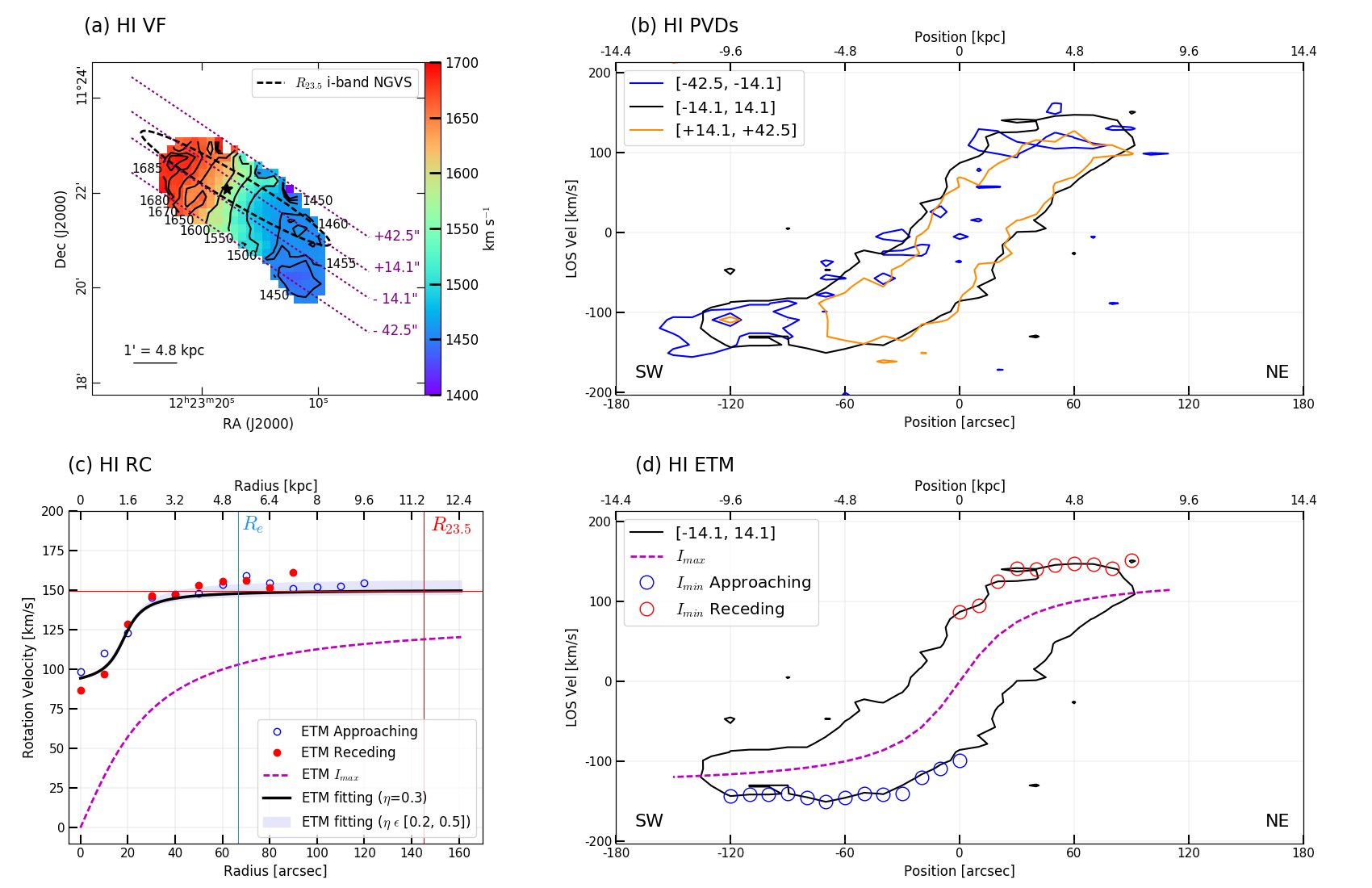}
    \caption{
HI emission data of NGC\,4330 from \citet{chung-2009}.
Panel (a): Cold gas velocity field on which are superimposed the positions of pseudo-slits ($\sim$28 arcsec wide) parallel to the kinematic major axis and the ellipse fitted to the isophote of surface brighness $\mu(i)=23.5\,\mathrm{\,mag\,arcsec^{-2}}$ of the NGVS $i$-band image, which traces the stellar disc. 
Panels (b) and (d): outermost contour of the position-velocity diagrams (PVDs). Panel (b) corresponds to the three pseudo-slits positions drawn in panel (a) and panel (d) corresponds to the central pseudo-slit position, the red and blue circles indicate the position of the minimum intensity of the receding and approaching side, respectively, while the dashed line indicates the maximum intensity.  Both entities are used to estimate the terminal velocity and to compute the RC using the ETM method. 
Panel (c): Rotation curve (RC) computed using the ETM, where the terminal velocity is represented by empty blue circles for the receding side and the filled red circles for the approaching side. The black solid line indicates the best fit to these points using  $\eta=0.3$, the shaded area shows the rotation velocity amplitude for $\eta= [0.2, 0.5]$, the purple dashed line the RC computed from the PVD maximum intensity, and the red and blue  vertical lines indicate the photometric and effective radius, respectively.
   }
   \label{HIPVD}
   
\end{figure*}
\begin{figure}[ht!]
\centering
\includegraphics[width=1\linewidth]{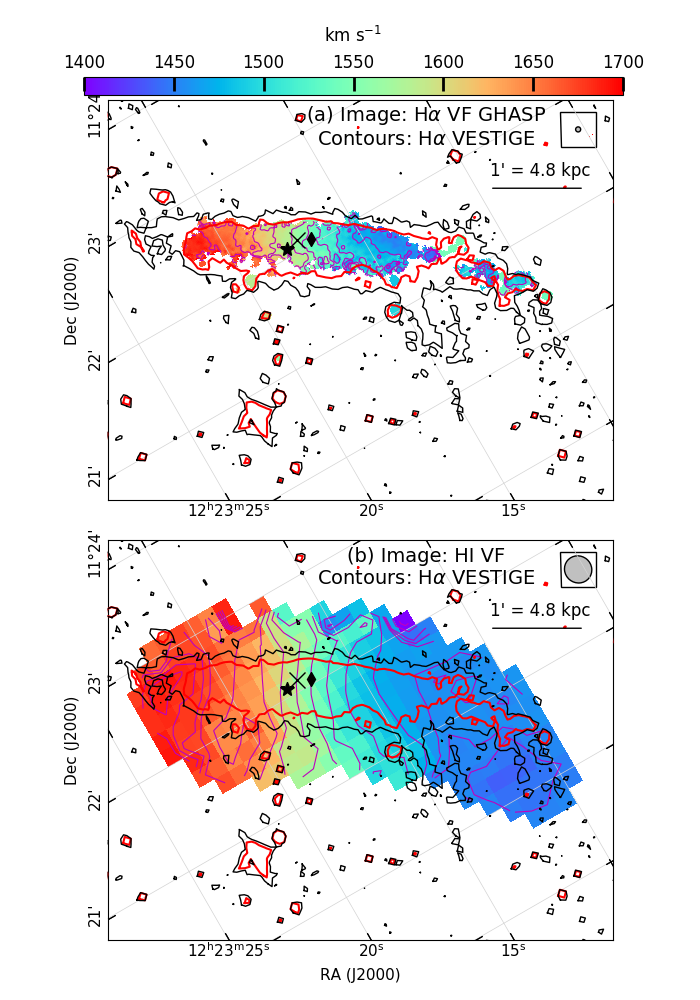}
    \caption{Comparison between the GHASP H$\alpha$ (upper panel) and the HI \citep[lower panel,][]{chung-2009} velocity fields of NGC\,4330.  The inner and outer black 
contours are at $\Sigma(H\alpha)$\,=\,3\,$\times$\,10$^{-18}$ and 1.25\,$\times$\,10$^{-16}\ \mathrm{erg\ s^{-1}\ cm^{-2}\, arcsec^{-2}}$, respectively, while 
the red contour $\Sigma(H\alpha)$\,=\,10$^{-17}\ \mathrm{erg\ s^{-1}\ cm^{-2}\, arcsec^{-2}}$.
The black diamond, the cross, and the star indicate the position of the CO, H$\alpha$, and HI emission kinematic centres, respectively. 
To align the maps horizontally, both are rotated $i=59^{\circ}$, the photometric $PA$ of the galaxy.
The grey ellipses encapsulated in a box located in the upper-right corner of each panel represent the sizes of the H$\alpha$ mean seeing disc (upper panel) and of the HI beam (upper panel).}
   \label{hiha}
\end{figure}

\begin{figure}[ht!]
\includegraphics[width=\linewidth]{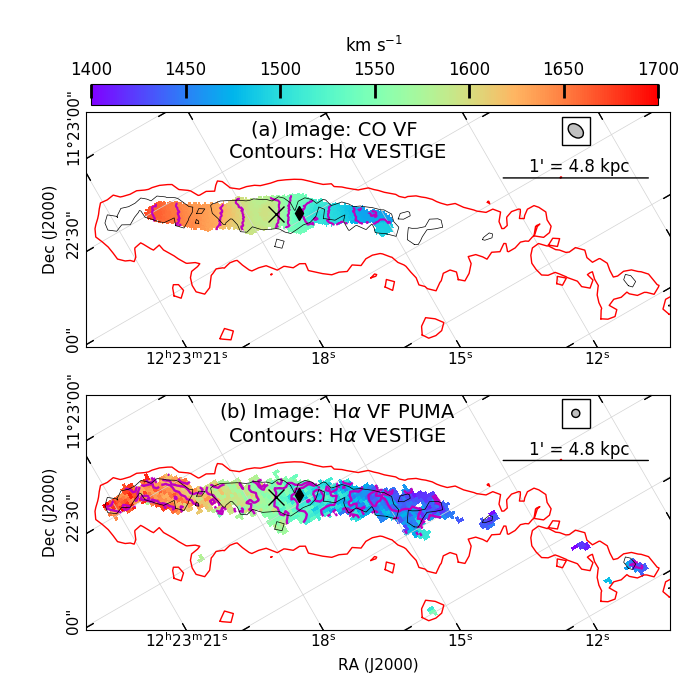}
\caption{Comparison between the  $^{12}\mathrm{CO}$ \citep[upper panel,][]{lee-2017} and the PUMA H$\alpha$ (lower panel) velocity fields of NGC\,4330. 
The inner black contours and the red outer contours are at 
$\Sigma(H\alpha)$\,=\,10$^{-17}$ and 1.25$\times$10$^{-16}\ \mathrm{erg\ s^{-1}\ cm^{-2}\, arcsec^{-2}}$, respectively.  
The black diamond and the cross indicate the position of the CO and H$\alpha$ emission kinematic centres, respectively.
To align the maps horizontally, both are rotated $i=59^{\circ}$, the photometric $PA$ of the galaxy.  
The grey ellipses located in the upper-right corner of each panel represent the sizes of the CO beam (upper panel) and of the H$\alpha$ mean seeing disc (lower panel). 
}
\label{COHA}
\end{figure}

\begin{figure}[ht!]
\centering
\includegraphics[width=\linewidth]{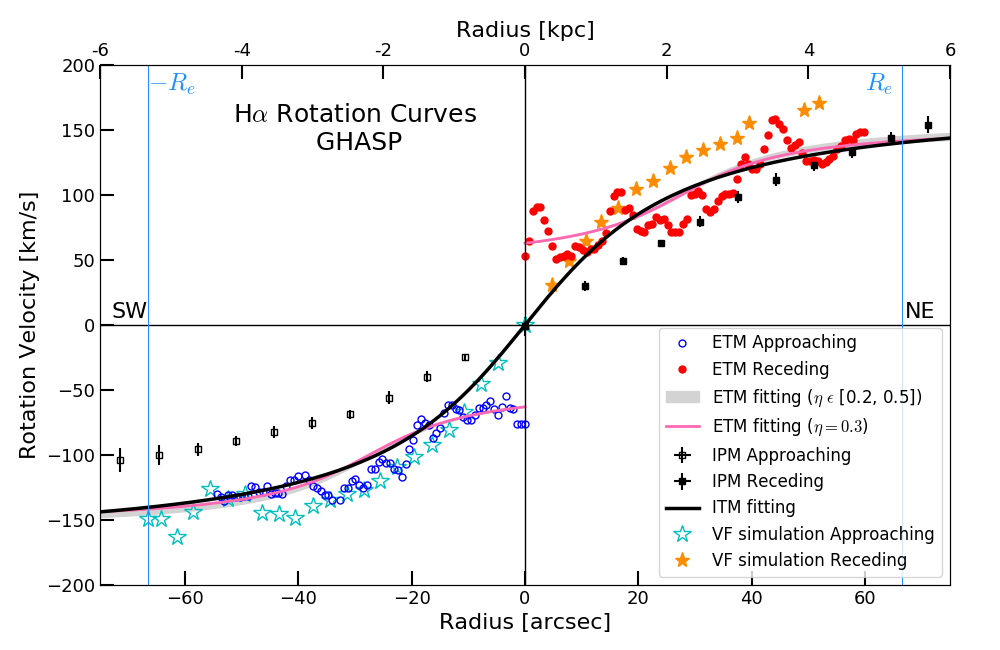}
\caption{Comparison between observed and simulated unfolded rotation curves.  The simulated velocity field shown in the panel (c) of Fig. \ref{n4330DataModelRV} has been used to measure the rotation curve along the major axis of the galaxy, the receding and approaching sides are respectively represented by filled orange and opened cyan stars ($\star$) and superimposed to the GHASP observed and modelled rotation curves already displayed in the panel (c) of Fig. \ref{RCET}, using the same symbols, colours and lines. The blue  vertical lines indicate the position of the effective radius, respectively.
}
\label{RC_sim_obs}
\end{figure}

\begin{figure}[ht!]
\centering
\includegraphics[width=9cm]{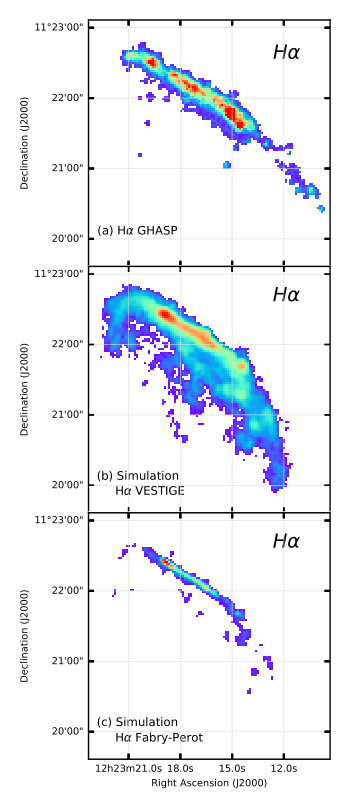}
\caption{Panel (a) GHASP and (panels (b) and (c)) simulated H$\alpha$ surface brightness of NGC\,4330, with scales identical to those used in Fig. \ref{Hacomposite}\,(d), 
i.e. with a threshold limit $\Sigma(H\alpha)$ $\simeq$  
10$^{-17}$\,erg\,s$^{-1}$\,cm$^{-2}$\,arcsec$^{-2}$.
The low level cuts used to display the simulated H$\alpha$ images have been chosen to show the low column density ionised gas reached by the simulations (panel (b)) and at higher surface brightness levels ($\Sigma(H\alpha)$ $\simeq$ 2$\times$10$^{-17}$\,erg\,s$^{-1}$\,cm$^{-2}$\,arcsec$^{-2}$, panel (c)), matching the typical surface brightness limit of VESTIGE and Fabry-Perot observations, respectively.
}
\label{n4330DataModelMono}
\end{figure}

\end{appendix}

\end{document}